\documentclass{aa}  

\usepackage{graphicx}
\usepackage{txfonts}

\usepackage{ulem}

\usepackage{xcolor}

\usepackage{soul}

\usepackage{float}

\usepackage{array, multirow, bigdelim, makecell, booktabs} 

\usepackage{threeparttable}

\usepackage{lscape}

\usepackage{longtable}

\usepackage{amssymb}

\usepackage[font={bf}]{subfig}
\usepackage[labelfont=bf]{caption}

\defcitealias{oh2017}{O+17}
\defcitealias{gagne2018a}{G+18}

\usepackage[colorlinks = true, linkcolor=blue, citecolor = blue]{hyperref}
\begin{document}

\title{The Volans-Carina association} 
\subtitle{An X-ray Bridge Between Younger and Older Stellar Populations}

   \author{Daniela Muñoz-Giraldo
          \inst{1}
          \and
          Beate Stelzer\inst{1}
          \and
          Alex Binks\inst{1}
          \and
          Enza Magaudda\inst{1}
          \and
          Stefanie Raetz\inst{1}
          \and
          Christian Schneider\inst{2}
          }

   \institute{Institut f\"ur Astronomie und Astrophysik, Eberhard Karls Universit\"at T\"ubingen, Sand 1, 72076 T\"ubingen, Germany \\ 
   \email{munoz-giraldo@astro.uni-tuebingen.de} 
    \and    
    Hamburger Sternwarte, Gojenbergsweg 112, 21029 Hamburg, Germany }

   \date{Received XX; accepted XX}

 
  \abstract
   {Nearby young moving groups (NYMGs) provide benchmarks for studying the evolution of magnetic activity and planetary environments at early ages. The Volans-Carina association (VCA), as a pre-main sequence association at a distance of $\sim$90\,pc, occupies a sparsely sampled region of parameter space between younger nearby associations and older distant open clusters.}
   {We reassess the census of the VCA and present the first nearly complete ($\sim$90$\%$) X-ray characterization of a NYMG.}
   {We have re-evaluated the membership probabilities of the VCA catalog presented in \cite{gagne2018a} with BANYAN $\Sigma$ as well as updated kinematics from {\it Gaia} DR3. Stellar parameters were derived from multi-wavelength spectral energy distributions, with the obtained bolometric luminosity and effective temperature being used in combination with pre-main sequence evolutionary models to obtain an isochronal age of the association. We joined data from the SRG (Spectrum Roentgen Gamma)/eROSITA and ROSAT all-sky surveys with a dedicated {\it XMM-Newton} Large Programme to construct a comprehensive X-ray catalog.}
   {Our updated census of the VCA comprises 29 highly probable members, 30 candidate members, and four uncertain objects. Isochronal analysis of the association yields an age of $80\pm20$\,Myr, consistent with previous estimates. We detect X-ray emission from 56 objects, reaching $\sim$90$\%$ completeness, one of the most complete X-ray censuses achieved for a NYMG. Low-mass members and candidates ($\lesssim0.6\,M_{\odot}$) have X-ray to bolometric luminosity ratios around the canonical saturation limit ($L_{\rm x}/L_{\rm bol}$ $\approx$ 10$^{-3}$). X-ray luminosity functions show the VCA to be consistent with the similarly aged Pleiades (in the F5 to M3 range) and significantly more active than the older Hyades (in the F5 to K4 range). The combination of an updated membership census and a nearly complete X-ray census establishes the VCA as a particularly robust calibration point for studies of the early evolution of coronal X-ray activity.}
   {}

   \keywords{stars: activity - stars: evolution – stars: magnetic field - stars: pre-main sequence – X-rays: stars}

   \maketitle
%

\section{Introduction}\label{sect:intro}

Nearby young moving groups \citep[NYMGs;][]{torres2000,zuckerman2001,torres2008,zuckerman2011} are gravitationally unbound, but co-moving, coeval stellar populations with ages $\lesssim 150\,$Myr and located within 100\,pc of the Sun. As such, NYMGs represent crucial benchmark populations for studying stellar magnetic evolution and planetary environments \citep{stauffer2016,carter2021}. Their proximity and low extinction enable detailed characterization, while their enhanced chromospheric and coronal emission \citep{kastner1997,mamajek1999,stelzer2000,torres2000,zuckerman2000,mamajek2002}, traced through UV and X-ray luminosities, distinguishes them from older field populations. Young stars are also expected to host forming or recently formed planetary systems whose atmospheres are exposed to intense X-ray and UV irradiation, making X-ray studies of NYMGs important for understanding atmospheric escape, magnetic star-planet interactions, and the early evolution of planetary habitability \citep{ribas2005,lammer2003,owen2012}.

The identification of NYMG members relies heavily on their shared kinematic properties. Modern surveys, particularly with {\it Gaia} \citep{gaiadr3}, have revolutionized the discovery of NYMGs by revealing the precise astrometric and kinematic fingerprints of nearby co-moving populations \citep[e.g.,][]{faherty2018,gagne2018a}. However, kinematics alone are insufficient to establish membership, as older field stars can share similar Galactic motions and contaminate NYMG samples. Independent youth diagnostics are therefore essential for distinguishing genuine young members from field interlopers. Among these diagnostics, X-ray emission has played a central role in the discovery and validation of NYMG members \citep[e.g.,][]{kastner1997,kastner2003,stelzer2000,stelzer2013,gagne2018a}, beginning with the ROSAT All-Sky Survey \citep[RASS][]{voges1999}, which provided the first systematic census of magnetically active young stars in the solar neighborhood. X-ray activity is a powerful tracer of stellar youth and magnetic evolution \citep{feigelson1999,stelzer2000,schneider2019}, with studies of X-ray luminosities and fractional X-ray luminosities ($L_{\rm x}/L_{\rm bol}$) in coeval populations providing important constraints from the pre-main sequence to the main sequence on coronal heating, the saturation of magnetic and stellar activity, and its mass dependence \citep{briceno1997,kastner1997,argiroffi2016,kastner2016}.

Previous X-ray studies of NYMGs, largely using ROSAT, were often limited by incomplete samples and sensitivity constraints, especially for low-mass stars \citep[e.g.,][]{stelzer2001,preibisch2005,richey2023}. The launch of the extended ROentgen Survey with an Imaging Telescope Array \citep[eROSITA;][]{predehl2021} on board the Spektrum-Roentgen-Gamma mission \citep[SRG;][]{sunyaev2021} now enables substantially more complete X-ray censuses of nearby young stellar populations, particularly intermediate-age associations where coronal activity levels are expected to be similar to very young populations, and significantly elevated relative to the field \citep{kastner2003,stelzer2013}. 

Within this context, the Volans-Carina association (VCA) is a particularly compelling target. First identified as a co-moving group by \citet[][herein, \citetalias{oh2017}]{oh2017} and subsequently expanded and characterized by \citet[][herein, \citetalias{gagne2018a}]{gagne2018a}, VCA has a distance of approximately 80–100 pc and an age of 89$^{+5}_{-7}$\,Myr, as estimated by \citetalias{gagne2018a}, slightly younger than the Pleiades \citep[112$\pm$5\,Myr;][]{dahm2015}. Its members, identified using the BANYAN (Bayesian Analysis for Nearby Young AssociatioNs) $\Sigma$ classification framework \citep{gagne2018b}, occupy a compact region of the sky close to the Galactic plane, likely contributing to its delayed recognition as a distinct association.

The VCA occupies an intermediate and underexplored region of the distance–age parameter space \citepalias{gagne2018a}. It is older than the youngest NYMGs typically targeted in X-ray studies, yet significantly closer and less populous than intensively-studied open clusters of similar age. This makes the VCA uniquely well suited for a complete X-ray census, as its sparse population permits study of almost all known members. As a result, the VCA offers an opportunity to bridge studies of very young associations and older clusters, providing new insight into how coronal emission evolves as stars approach the main sequence.

We reassess the VCA census of \citetalias{gagne2018a} in Sect.~\ref{sect:census}, recalculating kinematic membership probabilities using {\it Gaia} DR3 astrometry. In Sect.~\ref{sect:stellarParameter}, we derive effective temperatures and bolometric luminosities for VCA members and candidates using SED fits, comparing them with theoretical isochrones to obtain an independent age estimate of the VCA. We present the X-ray catalog of the VCA in Sect.~\ref{sect:Xrays} using eROSITA and RASS data, as well as a dedicated {\it XMM-Newton} Large Program. In Sect.~\ref{sect:discussion}, we discuss the implications of the nearly complete X-ray census of the VCA, including the coronal activity properties, comparisons with other young associations and clusters, and constraints on the age and activity evolution of the association. We present our conclusions in Sect.~\ref{sect:conclusions}.

\section{Census of the VCA}\label{sect:census}

\subsection{The initial target list from \cite{gagne2018a}}\label{subsect:startPoint}

The VCA was first identified by \citetalias{oh2017}, who reported eight co-moving stars in Volans–Carina based on astrometry from {\it Gaia} DR1 \citep{gaiadr1}. \citetalias{gagne2018a} used updated 3D positions and velocities from {\it Gaia} DR2 \citep{gaiadr2}, to identify 21 candidate members whose Galactic positions ({\it XYZ}) and space velocities ({\it UVW}) matched with the mean kinematics of the 8 members selected in \citetalias{oh2017}. A subsequent kinematic re-analysis which incorporated supplementary radial velocity (RV) measurements led to the inclusion of c\,Car, the highest mass member of the VCA (spectral type B8 II), whilst three of the \citetalias{oh2017} stars lacked RV measurements at the time and were removed from the \citetalias{gagne2018a} list.

This list of 19 kinematically-selected highly probable members (HPMs) formed the basis of an updated kinematic model incorporated in BANYAN $\Sigma$, which led to an additional 46 candidate members (CMs) which were found through two complementary searches designed by \citetalias{gagne2018a} to probe both the stellar and substellar populations. A {\it Gaia} DR2-based search within 125\,pc targeted previously unidentified stellar members, including sources lacking RV information, while a separate search using infrared surveys like 2MASS \citep{skrutskie2006} and AllWISE \citep{wright2010,kirkpatrick2014} focused on fainter objects not detected with {\it Gaia}, enabling the identification of potential substellar candidates.

Many of the HPMs and CMs identified in \citetalias{gagne2018a} require additional data to confirm their membership, either via a youth-based diagnostic or a RV measurement to test the kinematic commonality in 3-D (or both). If all of these 65 targets turn out to be members, \citetalias{gagne2018a} find that the VCA census is complete above a mass of $\sim$0.2M$_{\odot}$.

\subsection{A {\it Gaia} DR3-based updated census for the VCA}\label{subsect:updated_census}

In this section, we revisit the membership probabilities and status of the VCA catalog from \citetalias{gagne2018a} using updated astrometric data from {\it Gaia} DR3 \citep{gaiadr3}, which provides significantly improved precision on parallax and proper-motion measurements. Moreover, it approximately quadruples the number of sources with RV values (see Sect.~\ref{subsubsect:RVs}). Table~\ref{table:vcaCatalog} presents the catalog of HPMs and CMs of the VCA, along with the corresponding VCA IDs assigned by \citetalias{gagne2018a}, which are adopted throughout this paper when referring to individual stars.

\subsubsection{Astrometric data}\label{subsubsect:astrometric_data}

All but one HPMs and CMs of the VCA from \citetalias{gagne2018a} have {\it Gaia} DR3 counterparts with 5-parameter astrometry, photometry and distances reported by \cite{bailer2021}, which will be used throughout this study and are provided in col. 3 of Table~\ref{table:vcaCatalog}. The only exception is VCA~48, which lacks {\it Gaia} $G_{\rm BP}$ and $G_{\rm RP}$ photometry, and is therefore not included in the majority of our figures.

Three CMs (VCA~47, VCA~54, and VCA~43) have distances between $\sim$400\,pc and $\sim$9000\,pc, clearly inconsistent with the mean VCA distance of $86\pm5$\,pc \citepalias{gagne2018a}. These discrepancies suggest that these systems are unlikely to be genuine members, and we exclude them from the list of VCA candidates and from all subsequent analyses.

\subsubsection{A new common proper-motion binary component}\label{subsubsect:CPMbinary}
With the aim of identifying possible optically resolved common proper motion (CPM) binaries we searched for additional {\it Gaia} DR3 sources within 30" of each VCA object (corresponding to a physical separation of $\sim2.5\times10^{3}\,$AU at 80\,pc) whose parallaxes and proper motions agree within 3$\sigma$. This analysis revealed three resolved CPM binaries within the VCA catalog (see Table~\ref{table:vcaCatalog}). One additional resolved CPM binary was identified composed of a HPM (VCA~15) and a star not included in the \citetalias{gagne2018a} candidate list for the VCA (2MASS~J10223677-5848037). We tentatively add 2MASS~J10223677-5848037 as a new VCA candidate.

\subsubsection{Radial velocities}\label{subsubsect:RVs}

\begin{figure}
    \centering
    \includegraphics[width=\columnwidth,trim=0.4cm 0.3cm 0.3cm 0.3cm,clip]{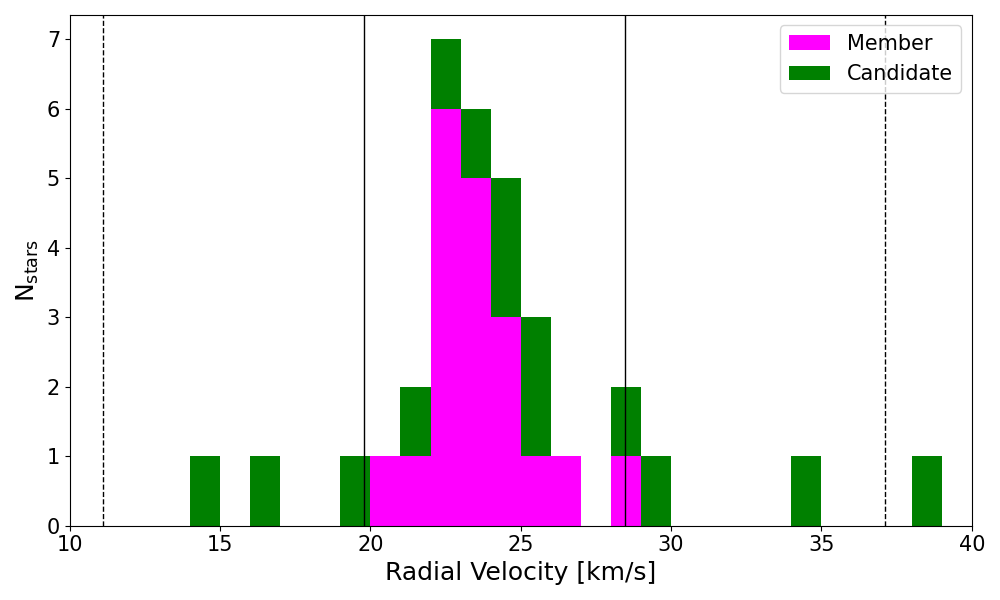}
    \caption{Distribution of RVs retrieved for $19$ HPMs and $14$ CMs of the VCA. We show the 1$\sigma$ (solid line) and 3$\sigma$ (dashed line) ranges.}
    \label{fig:rvVCA}
\end{figure}

A complete 3-D kinematic membership test requires a RV measurement. If a target has an RV measurement, we incorporate it into our tests defined in Sect.~\ref{subsubsect:kin_test}. Following \citetalias{gagne2018a}, we retain only RV measurements with error bars $<10\,$km/s. 

We first searched for RV values using {\it Gaia} DR3, finding 36 stars in our sample with a reported RV measurement, 29 of which satisfy the aforementioned RV criterion. To supplement the {\it Gaia} DR3 RV data, we performed an automated search for RV measurements using a bespoke software program available on GitHub\footnote{\protect\url{https://github.com/alexbinks/RV\_selection}}, which queries the VizieR database \citep{vizier2000} using a search radius of 5'' on the J2000 coordinates. To avoid duplication, the code retains only RV values from the original catalog. For stars with multiple RVs, the final RV (and error bar) is selected as the weighted average, defined by the square of the RV uncertainties. Supplementary RVs are available for 16 stars, where 12 of these are RVs from {\it Gaia} DR2 (all of which also have a RV from {\it Gaia} DR3), and 4 are RVs not based on {\it Gaia}, bringing the total number of stars with an RV measurement to 33; comprising all 19 HPMs and an additional 14 CMs (which includes the CPM companion to VCA 15). The RV adopted for each star, and their literature reference sources are listed in cols. 4 and 5 of Table~\ref{table:vcaCatalog}, respectively.

The resulting RV distribution is shown in Fig.~\ref{fig:rvVCA}, with a mean value of $24.1\pm4.3$\,km\,s$^{-1}$. As expected for a compact association such as VCA, the RV dispersion is small compared to e.g. the typical dispersion of 10\,km\,s$^{-1}$ for the Hyades and 7\,km\,s$^{-1}$ for the Pleiades, AB Dor, and UMa \citep{chen1997,chereul1999}. While the mean RV of VCA is not unique among nearby young associations, as several NYMGs have mean RVs consistent with the VCA value within uncertainties \citep{gagne2018b}, the narrow distribution observed here provides an additional indication of the kinematic coherence of the association. All but one CM have RVs within $3\,\sigma$ of the mean value, but even the RV of this outlier candidate is consistent with $3\,\sigma$ of the mean of the VCA within its large individual uncertainty (4.7\,km\,s$^{-1}$). For the 12 objects with multiple RV measurements, no evidence is found for short-period multiplicity that could significantly bias the kinematic analysis.

\subsubsection{Kinematic membership selection tests}\label{subsubsect:kin_test}

We re-assess the membership status for each of the 63 members and candidates using the BANYAN $\Sigma$ code. For the 33 stars with a suitable RV measurement (19 HPMs and 14 CMs) we run BANYAN $\Sigma$ where the inputs are the {\it Gaia} DR3 5-parameter data and the mean RV. In order to maintain or gain a classification as a HPM, stars are required to present; (1) a BANYAN $\Sigma$ membership probability $P_{\rm mem}\,>\,90\%$, and (2) spatial ($XYZ$) and velocity ($UVW$) coordinates lying within 15\,pc and 4\,km\,s$^{-1}$, respectively, of the mean values derived from the $19$ original members identified by \citetalias{gagne2018a}. Stars that satisfy the second kinematic criterion but have $P_{\rm mem} < 90\%$ are classified as uncertain members or candidates, depending on whether they were originally identified as HPMs or CMs.

For 29 of the 33 targets with RV measurements ($16$ HPMs and $13$ CMs) we find $P_{\rm mem}\,>\,90\%$. The four targets that failed this test still satisfy the second kinematic criterion. Among these four, there are 2 HPMs (VCA~1, and VCA~15) and 1 CM (the newly-identified CPM candidate 2MASS~J10223677-5848037) with $P_{\rm mem} \sim 50\%$, and one HPM (VCA~6) with $P_{\rm mem}<10\%$. We find that VCA~6 would qualify as a member if the RV ($=28\pm0.6$\,km\,s$^{-1}$) had not been used. However, we still regard it as a potential member because the RV is approximately 1$\sigma$ from the mean RV calculated in Sect.~\ref{subsubsect:RVs}.

The 30 remaining CMs that lack an RV measurement are found with $P_{\rm mem}>90\%$, which were calculated by marginalizing over all possible RV values. We maintain these as CMs, since their membership status awaits confirmation through an RV measurement. The BANYAN $\Sigma$ code returns an "optimal" RV for these targets, which is the value needed to ensure the $UVW$ vectors are co-moving with the VCA. These "optimal" RVs are displayed in col. 4 of Table~\ref{table:vcaCatalog}~for all CMs. 

\subsubsection{An updated catalog for the VCA}\label{subsect:VCAcatalog}

\begin{figure}
    \centering
    \includegraphics[width=\columnwidth,trim=0.4cm 0.1cm 1.5cm 1.4cm,clip]{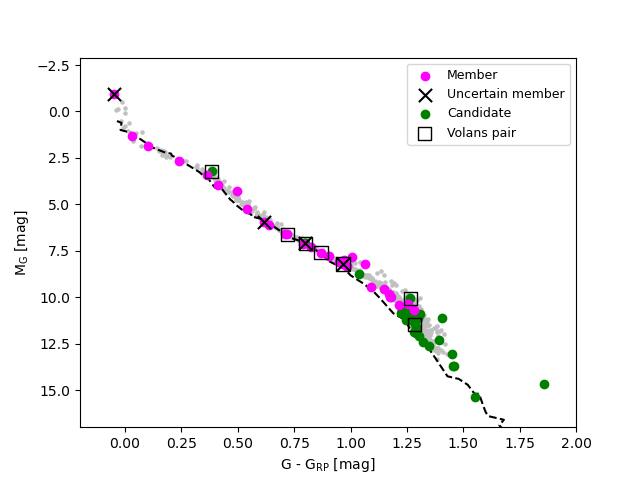}
    \caption{CMD for the VCA. CPM binaries are highlighted with black squares. As reference we show members of the Pleiades cluster \citep{alfonso2023} as grey dots and the \cite{pecaut2013} main-sequence.}
    \label{fig:cmdVCA}
\end{figure}

The updated census of the VCA is presented in Table~\ref{table:vcaCatalog}, and illustrated in the {\it Gaia} color-magnitude diagram (CMD) shown in Fig.~\ref{fig:cmdVCA}. The final sample comprises 29 members ($16$ previous HPMs and $13$ newly confirmed), 30 CMs, three uncertain members and one uncertain candidate. Three resolved CPM binaries are identified among the HPMs and CMs (indicated as black squares in Fig.~\ref{fig:cmdVCA}), as well as one additional resolved CPM binary involving an uncertain member (VCA~15) and an uncertain candidate (2MASS~J10223677-5848037).

\section{Characterizing the VCA}\label{sect:stellarParameter}

In this section, we describe how the stellar parameters of all VCA stars were computed, and how we use them to obtain an age estimate for the association. 

\subsection{Bolometric luminosities and effective temperatures}

The bolometric luminosity ($L_{\rm bol}$) is a fundamental parameter in studies of stellar activity, as it enters directly into the definition of the fractional X-ray luminosity, $L_{\rm x}/L_{\rm bol}$. While X-ray luminosity is known to depend on stellar mass \citep[see e.g.][]{preibisch2005}, normalization by $L_{\rm bol}$ largely mitigates this dependence, enabling meaningful comparisons of activity levels across the full mass range of a stellar association or moving group. An accurate determination of $L_{\rm bol}$ is, therefore, essential for placing the X-ray properties of the VCA into the broader context of stellar activity evolution.

Initial estimates of $L_{\rm bol}$ can be obtained from the positions in the CMD (Fig.~\ref{fig:cmdVCA}). However, for the late-type stars that dominate this VCA sample, redder optical and infrared wavelengths provide a more reliable characterization of the stellar photospheres. We therefore determine stellar parameters using multi-band photometry taken from publicly available surveys covering a wide range of wavelengths from ultraviolet to infrared, specifically GALEX \citep{martin2005}, {\it Gaia} DR3, 2MASS, and WISE. Spectral energy distributions (SEDs) are constructed by cross-matching counterparts across the different catalogs with an appropriate radius for each survey \citep{munoz2024}. The resulting SEDs are fitted using the Virtual Observatory SED Analyzer \citep[VOSA;][]{bayo2008}, and bolometric luminosities and effective temperatures ($T_{\rm eff}$) are derived. 
SEDs were constructed for all 63 stars of the VCA.

GALEX counterparts were identified for three HPMs (VCA~6, VCA~10, and VCA~18) and one CM (VCA~29), with angular separations smaller than 1.6\arcsec, indicating reliable associations. These results are consistent with \citetalias{gagne2018a}, who reported GALEX detections for the same four objects in the VCA. Overall, and in agreement with previous work, ultraviolet detections among candidates and members remain rare.

The SEDs were fitted with VOSA using a late-type photosphere model with solar metallicity \citep[BT-Settl;][]{allard2003}, allowing the effective temperature to vary freely. The resulting $T_{\rm eff}$ values are consistent with the expectations based on the spectral type estimates reported by \citetalias{gagne2018a} (see Table~\ref{table:vcaCatalog}). 

\subsection{Isochronal age estimation of the VCA}\label{subsect:isochrones}

\begin{figure}
    \centering
    \includegraphics[width=0.49\textwidth,trim=0.7cm 0.2cm 1.2cm 1.4cm,clip]{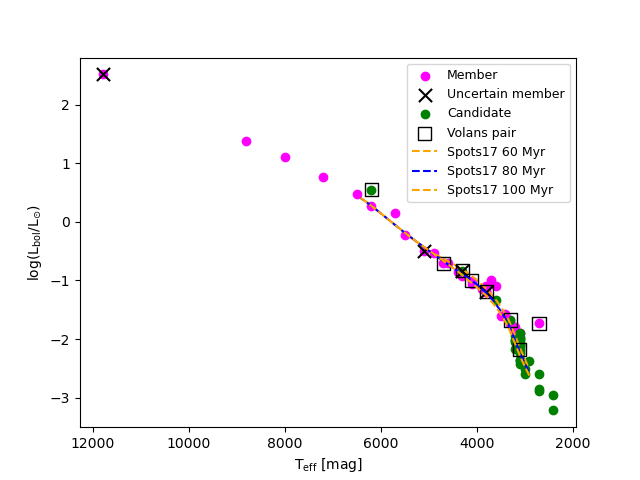}\hfill
    \includegraphics[width=0.49\textwidth,trim=0.4cm 0.2cm 1.2cm 1.4cm,clip]{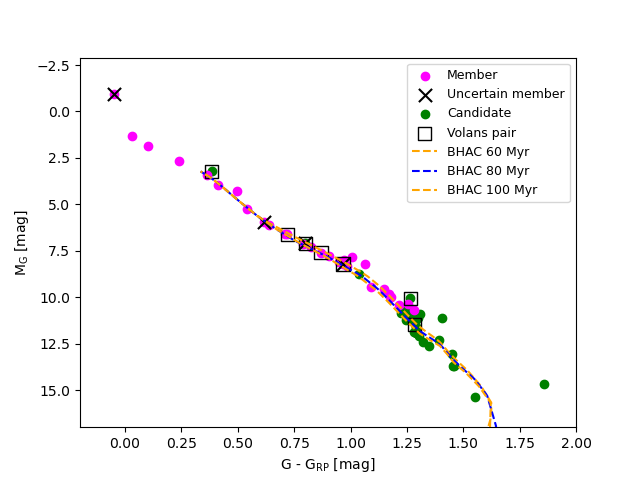}
    \caption{Age estimation of the VCA showing the preferred age for the VCA (85\,Myr; blue isochrone) with the age range (orange isochrones) obtained from the best fitting isochrones. \textit{Upper panel:} HRD. \textit{Lower panel:} CMD.}
    \label{fig:HRfit}
\end{figure}

We use the position of the VCA stars in the {\it Gaia} CMD and in the Hertzsprung-Russell diagram (HRD) to estimate the age of the association from comparison to theoretical isochrones. The HRD 
for the VCA was constructed using the $(L_{\rm bol}, T_{\rm eff})$ pairs resulting from the SED fittings. The two diagrams (shown in Fig.~\ref{fig:HRfit}) provide independent age estimates that we then combine into our final range of plausible association ages. 

To estimate the isochronal age, we adopt two complementary fitting approaches. In the first method, all stars are fitted simultaneously to isochrones from different evolutionary calculations (see below). We initially consider only single stars, excluding resolved CPM binaries, and subsequently restrict the sample to M stars that are not part of resolved or unresolved multiple systems. This restriction to only M stars is motivated by the increased sensitivity of low-mass stars to age differences at the expected age of the VCA ($\sim$89\,Myr; \citetalias{gagne2018a}), where isochrones are more widely separated in the low-temperature regime. In the second method, isochrones are fitted individually to each star, and the resulting age estimates are combined to derive a representative value for the association. This latter approach is applied exclusively to the subset of single M stars. The advantage of the star-by-star fitting is that an uncertainty on the age of the whole group is naturally obtained from the spread of the individual ages. In both methods, the goodness of fit is evaluated through a $\chi^2$-statistic, and the preferred solution is identified as the isochrone with the reduced $\chi^2$-value closest to unity.

For the CMD analysis, we consider isochrones from the PARSEC \citep{nguyen2022,nguyen2025}, BHAC \citep{baraffe2015}, MIST \citep{choi2016,dotter2016}, and BaSTI \citep{hidalgo2018} families. SPOTS \citep{somers2020} isochrones are not included in this case, as they do not adequately sample the low-mass regime at the relevant ages. Using the first method, the best-fitting solution corresponds to a 60\,Myr BHAC isochrone, when considering all single stars, and a 80\,Myr BHAC isochrone when restricting the sample to single M stars. Applying the second method to the single M star subset, and using only BHAC isochrones, as motivated by the first method, yields an average age of 93$\pm$37\,Myr.  

For the HRD analysis, we consider in addition to the four above mentioned model families also SPOTS isochrones with spot coverage fractions of 17$\%$, 34$\%$, and 51$\%$. In this case, the first method yields best-fit ages of 70\,Myr for a SPOTS isochrone with 17$\%$ spot coverage when considering all single stars as well as when restricting the sample to single M stars. Applying the second method to the single M star subset, the preferred SPOTS model (17$\%$) results in an average age of 62$\pm$11\,Myr.

Taken together, the CMD and HRD analyses yield consistent results within uncertainties, indicating an age range of approximately 60–100\,Myr for the VCA. This range encompasses the best-fit solutions from both fitting approaches and reflects both systematic differences between isochrone families and intrinsic scatter among individual stars. In the following sections, we adopt this interval as a conservative estimate of the isochronal age of the association with a preferred age of 80\,Myr. Fig.~\ref{fig:HRfit} illustrates, overplotted on the data, representative isochrones corresponding to the lower and upper bounds of this age range, namely 60–100\,Myr for BHAC isochrones in the CMD and for SPOTS models with 17$\%$ spot coverage in the HRD. Our conservative age for the VCA of 80$\pm$20\,Myr is in agreement with the 89$_{-7}^{+5}$\,Myr age derived by \citetalias{gagne2018a} using {\it Gaia} DR2. 

\section{The VCA in X-rays}\label{sect:Xrays}

In this section, we present a comprehensive X-ray census of the VCA, constructed from data from the eROSITA and ROSAT all-sky surveys, and a dedicated {\it XMM-Newton} Large Programme (Prop-ID 094096; PI Stelzer). The resulting catalog compiling all X-ray data achieves a detection completeness exceeding 90$\%$, representing one of the most complete X-ray samples assembled to date for a nearby young association. 

\subsection{eROSITA detections}\label{subsection:eROSITA}

Between 2019 and 2021, eROSITA completed four successive scans of the entire sky each lasting six months, referred to as eRASS\,1 to eRASS\,4, and a fifth partial scan named eRASS\,5 \citep{merloni2024}. The corresponding source catalogs are generated at the Max Planck Institut für extraterrestrische Physik (MPE) in Garching, Germany, using the eROSITA Science Analysis Software System \citep[eSASS version 240410.0.4;][]{brunner2022}. These catalogs include all sources detected in the hemisphere of the sky with German data rights, defined in Galactic coordinates by longitudes $l\,>\,179.94^\circ$. All HPMs and CMs of the VCA are located in the area with German data rights.

In order to maximize sensitivity to the expected low X-ray luminosities of the faintest VCA members, we base our analysis on the combined eRASS:5 catalog, which sums the data from all four individual all-sky surveys and the fifth partial one. The eRASS:5 catalog employed here was produced using the latest data processing version 030\footnote{The source catalog used in our work is\\   all{\textunderscore}s4{\textunderscore}s5{\textunderscore}SourceCat1B{\textunderscore}c030{\textunderscore}240905{\textunderscore}poscorr{\textunderscore}mpe{\textunderscore}photom.fits.}. It contains data in a single energy band from $0.2$–$2.3$ keV.

\begin{figure}
    \centering
    \includegraphics[width=\columnwidth,trim=0.4cm 0.3cm 0.3cm 0.3cm,clip]{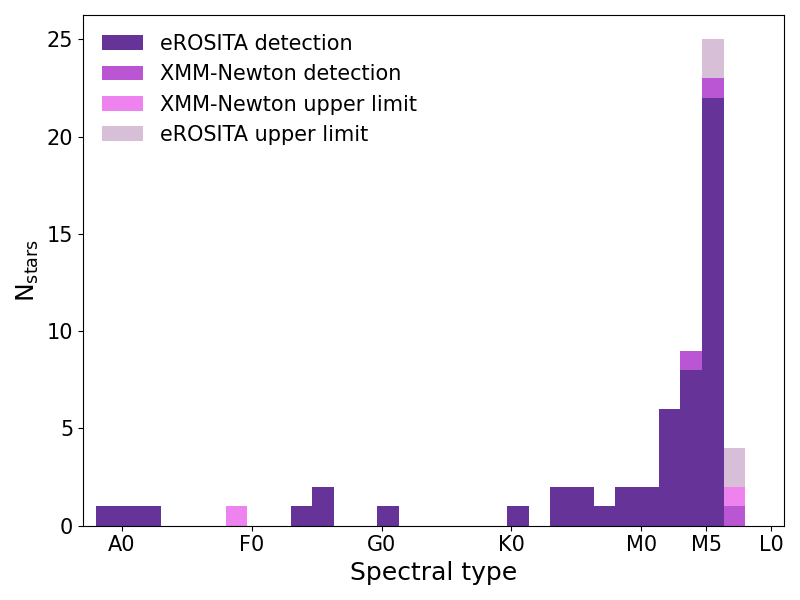}
    \caption{Distribution of spectral type of the VCA members and candidates with an X-ray detection or upper limit from eROSITA or {\it XMM-Newton}.}
    \label{fig:spteROSITA}
\end{figure}

The positions of the VCA stars from our updated catalog were first corrected to the mean epoch of the eRASS:5 observations by using their individual {\it Gaia} DR3 proper motions. The corrected coordinates were then cross-matched with the eRASS:5 source catalog. All possible X-ray counterparts were checked for an optical-to-X-ray separation (sep$_{\rm ox}$) that satisfies sep$_{\rm ox} \leq 30^{\prime\prime}$, and that is smaller than five times the 1$\sigma$ positional uncertainty of eROSITA ($POS\_ERR$), that is  sep$_{\rm ox}\,<\,5\,\times POS\_ERR$. In addition, we require the matched eROSITA sources to be consistent with point-like emission, as indicated by an extension flag of $EXT\,=\,0$ in the eRASS catalog. Through this methodology we find that 27 of the 29 HPMs, 23 of the 30 CMs, and all three uncertain members have an eROSITA counterpart within 10" (see Table~\ref{table:vcaCatalog}). 

To verify that the VCA stars are effectively responsible for the X-ray emission we followed the methodology of \cite{magaudda2022}, performing a ``reverse match" between the eRASS:5 X-ray source and the \textit{Gaia} DR3 catalog. For this we produce a list of all \textit{Gaia} sources within $60^{\prime\prime}$ of the eRASS source. The coordinates of these {\it Gaia} sources were then corrected using their individual proper motions to the mean epoch of the eRASS:5 catalog. Using the corrected coordinates, we produce a final match of {\it Gaia} sources within $30^{\prime\prime}$ of the eRASS source. As a result of this procedure, we find that all VCA objects, that are not in a CPM binary, are the closest \textit{Gaia} DR3 source to both the eRASS:5 source and to the Earth. We, therefore, consider in all cases the VCA object to be the correct optical counterpart to the X-ray source. In the cases of the optically resolved CPM binaries, both components of the binary are matched to the same eRASS:5 source, such that these objects need special treatment when it comes to an evaluation of their X-ray properties (see below). 

In order to allow for a comparison across instruments, as well as with previous X-ray studies, we calculated the X-ray flux in the ROSAT band (0.1-2.4\,keV). For this we used the count rate reported in the eRASS:5 catalog and a conversion factor ($CF$) defined as $CF\,=\,Flux/CountRate$. The values we adopted are dependent on the spectral type with $CF\,=\,1.06\times10^{-12}$\,erg\,cm$^{-2}$\,cts$^{-1}$ for FGK stars \citep{bennedik2026} and $CF\,=\,8.78\times10^{-13}$\,erg\,cm$^{-2}$\,cts$^{-1}$ for M stars \citep{caramazza2023}. We combined the X-ray fluxes with the distances to calculate the X-ray luminosity of each object. For the CPM binaries we computed the individual X-ray luminosities by dividing the observed value by the number of components in the stellar system \citep[see e.g.][]{stelzer2001}. The fractional X-ray luminosities, $L_{\rm x}/L_{\rm bol}$, were calculated using the above X-ray luminosities and the bolometric luminosities derived in Sect~\ref{sect:stellarParameter}. In the case of CPM binaries this means that the components share the same X-ray luminosity but different $L_{\rm x}/L_{\rm bol}$ values.

With eROSITA alone we reach a completeness of $\sim$85$\%$ in the X-ray census of the VCA. The distribution of X-ray detections for different spectral types is shown in Fig.~\ref{fig:spteROSITA}, with the latest-type star detected with eROSITA being an M5 star. As expected, stars with spectral type late-M and beyond remain undetected due to the eROSITA sensitivity limit. This non-detection of several late-type stars, as well as one earlier-type star, motivated targeted follow-up observations with {\it XMM-Newton} aimed at improving the completeness of the X-ray census.

\subsection{XMM-Newton detections}

The \textit{XMM-Newton} observations of $14$ HPMs and CMs of the VCA that were not detected with  eROSITA at the time of the Large Programme proposal (using the eRASS:3 catalog) were performed with all EPIC instruments \citep{struder2001,turner} in \texttt{FULL FRAME MODE} using the \texttt{THIN} filter. The summary of these observations is shown in Table~\ref{table:xmmObservations}.

\begin{table}
    \begin{threeparttable}
    \centering
    \caption{{\it XMM-Newton} observation log.}             
    \label{table:xmmObservations}  
    \begin{tabular}{c cccc }
    \hline              
    \noalign{\smallskip}
        VCA  & Obs. ID & Obs. Date & Total Exp. & Flag \\
        ID  & & & [ks] & \\
        \noalign{\smallskip}
        \hline              
        \noalign{\smallskip}
        22  & 0940960101 & 2025-03-06 & 21.0 & 3 \\
        26  & 0940960201 & 2024-08-15 & 31.8 & 3\\
        34  & 0940960301 & 2024-10-31 & 21.3 & 1\\
        35  & 0940960401 & 2024-11-23 & 59.6 & 3 \\
        37  & 0940960501 & 2025-01-11 & 22.1 & 1 \\
        38  & 0940960601 & 2024-11-24 & 16.7 &\\
        41  & 0940960701 & 2024-05-10 & 10.8 & 1 \\
        44  & 0940960801 & 2025-01-26 & 20.4 &\\
        48  & 0940960901 & 2024-05-04 & 10.7 &\\
        50  & 0940961001 & 2024-12-13 & 32.8 & 2\\
        53  & 0940961101 & 2024-08-10 & 10.7 &\\
        54  & 0940961201 & 2024-05-08 & 44.6 & 2\\
        56  & 0940961301 & 2024-08-01 & 14.4 & 2\\
        57  & 0940961401 & 2024-07-29 & 10.7 &\\
        \noalign{\smallskip}
        \hline 
    \end{tabular}
    \begin{tablenotes}
        \small
        \item Flag: 1 for observations where the {\it XMM-Newton} instruments experienced "Full Scientific buffer" during the exposure, 2 for observations with elevated particle background and a source non-detection, and 3 for observations with elevated particle background and a source detection.
    \end{tablenotes}
    \end{threeparttable}
\end{table}

Of the $14$ HPMs and CMs observed with {\it XMM-Newton}, three had unusable data due to the EPIC/pn instrument experiencing "Full Scientific buffer" during the exposure preventing an analysis. The remaining 11 datasets were processed further, with data reduction carried out using the {\it XMM-Newton} Science Analysis System (SAS; version 22.0.0).  

To mitigate the effects of variable particle background, we screened the event lists of the observations by retaining only time periods below an appropriate count rate threshold selected individually for each object, maximizing the sensitivity to faint emission. Background conditions vary significantly from observation to observation. Five datasets are only mildly affected by soft-proton flaring, while six display elevated particle background. 

For the cleaned exposures, we applied standard filtering criteria, keeping single and double events (\texttt{PATTERN} $\leq$ 4), requiring \texttt{FLAG}\,=\,0, and restricting the analysis to the calibrated events channels (200 $\leq$ PI $\leq$ 12000). Source detection was then performed following a customized procedure built upon the steps implemented in the SAS task \texttt{EDETECT\textunderscore CHAIN} for three energy bands, 0.2–1.0 keV (soft), 1.0–2.0 keV (medium), and 2.0–12.0 keV (hard). The optical coordinates, propagated with the {\it Gaia} DR3 proper motions to the {\it XMM-Newton} observing date, are matched with the X-ray catalog produced during source detection in order to identify the correct X-ray counterpart.

For three of the observations displaying elevated particle background, the enhanced background combined with the intrinsically low source flux results in non-detections. One of these objects is VCA~54, which we reject as part of the VCA following our kinematic analysis presented in Sect.~\ref{subsect:startPoint}. Therefore, we derive upper limits for only two VCA objects, VCA~50 and VCA~56 (see Sect.~\ref{subsect:upperLimits}).

A circular photon extraction region with radius of $30^{\prime\prime}$ was defined for the spectral and temporal analysis, which was centered on each EPIC/pn source position. Similarly, for the background we chose an adjacent circular region with a radius of $30^{\prime\prime}$, making sure it did not include any other X-ray sources.  

We present here a general overview and highlights from our analysis of the {\it XMM-Newton} observations of only the eight detected VCA stars. The addition of three CMs that are not detected with eROSITA increases the completeness of the X-ray census of the VCA to $\sim$90$\%$ (see Fig.~\ref{fig:spteROSITA}). This level may reach $\sim$95$\%$ once the three stars affected by "Full Scientific buffer" are re-observed. At present, only six stars remain without X-ray detections: the three late-type CMs (M5–M7) for which the {\it XMM-Newton} observations were compromised due to the instruments experiencing "Full Scientific buffer", two stars (A9 and M8) with {\it XMM-Newton} non-detections, and one M8 candidate that was not targeted with {\it XMM-Newton} due to its expected faintness and that also remained undetected with eROSITA. For these six objects we derived upper limits from the most sensitive available X-ray observation, that is {\it XMM-Newton} for the former two groups and eROSITA for the latter one (as explained in Sect.~\ref{subsect:upperLimits}).

\subsubsection{X-ray light curves}

\begin{figure*}
    \centering
    \parbox{\textwidth}{\includegraphics[width=0.47\textwidth,trim=0.2cm 0.3cm 0.1cm 0.2cm,clip]{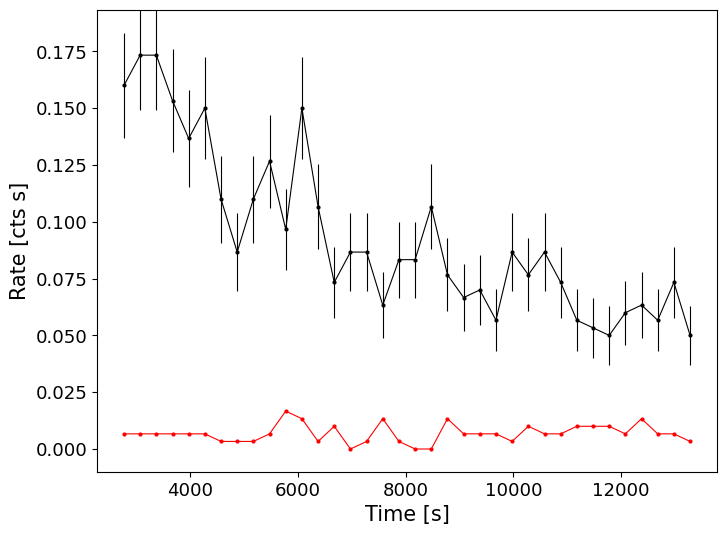}\hfill
    \includegraphics[width=0.47\textwidth,trim=0.2cm 0.2cm 0.1cm 0.97cm,clip]{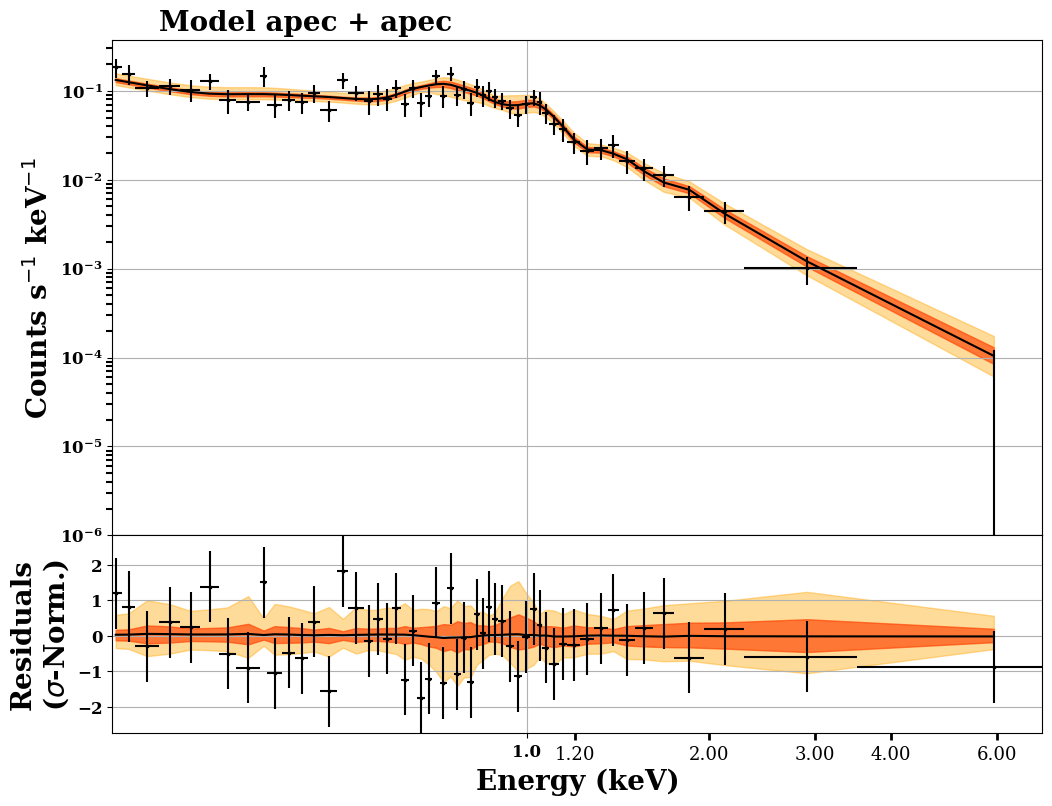}}\\
    \parbox{0.49\textwidth}{\includegraphics[width=0.47\textwidth,trim=0.3cm 0.2cm 0.2cm 0.2cm,clip]{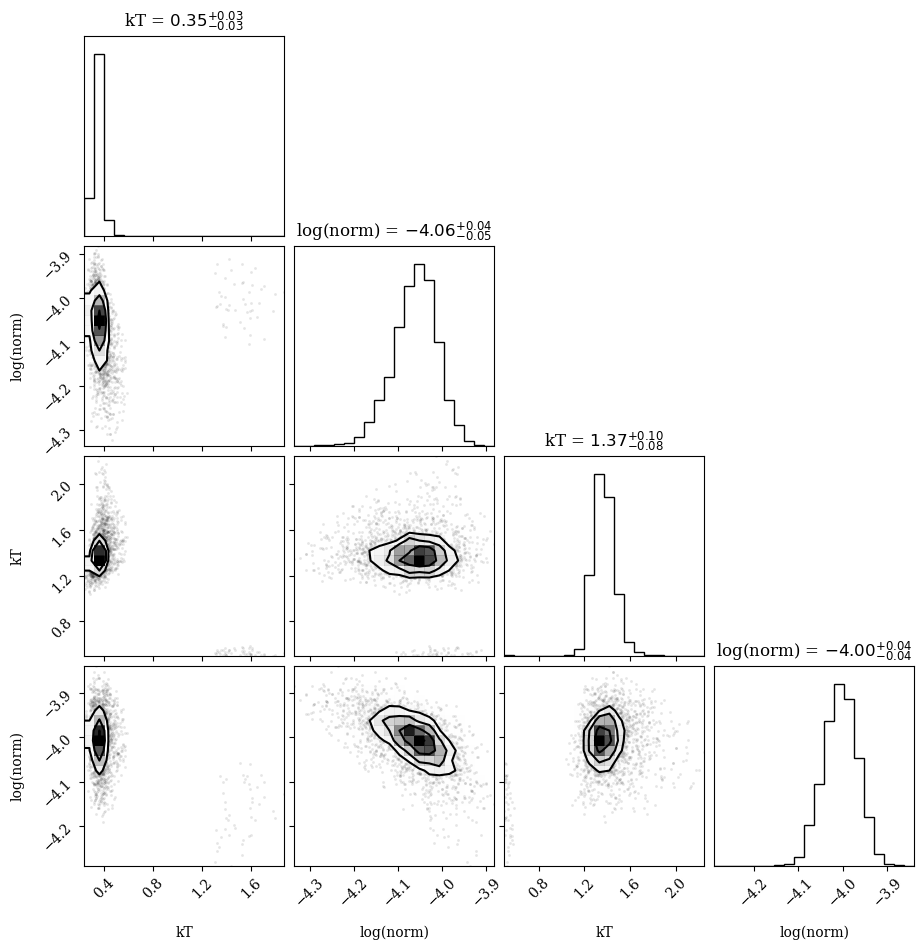}}\hfill
    \parbox{0.49\textwidth}{\caption{X-ray light curve, spectrum, and best-fit spectral parameters of the brightest X-ray source, VCA~48. {\it Upper left panel:} EPIC/pn light curve in the {\it XMM-Newton} broadband (0.2 -12 keV) with a bin size of 300\,s. The X-ray light curve represents the background-subtracted source signal (black), with the background signal (red) included for comparison. {\it Upper right panel:} EPIC/pn spectrum with best-fit two-component {\texttt APEC} model, and residuals. The 1$\sigma$ and 3$\sigma$ confidence regions are shown in orange and light orange respectively. {\it Lower left panel:} Corner plots for the best-fit spectral parameters, showing contours for the 1$\sigma$, 2$\sigma$, and 3$\sigma$ confidence regions.
    }\label{fig:1}}
\end{figure*}

We extracted background-subtracted EPIC/pn light curves in the {\it XMM-Newton} broad-band (0.2 -12.0 keV) for all 8 VCA objects detected in the {\it XMM-Newton} observations using the SAS task \texttt{EPICLCCORR}, which also corrects for instrumental effects, on an event list that was previously barycenter corrected using the SAS tool \texttt{BARYCEN}. For six of the 8 objects, the light curves display relatively stable emission at low count rate throughout the duration of the observations, with no clear evidence of variability (see figures in Appendix~\ref{appen:xmmSpectraAll}).

In contrast, two sources (VCA~48, and VCA~57) exhibit pronounced variability consistent with flaring events. In the case of VCA~57 (see Fig.~\ref{fig:xmmSpectraApend2}), the count rate increases considerably relative to the quiescent level, lasting $\sim$30$\%$ of the observation. For VCA~48, shown in Fig.~\ref{fig:1}, the light curve clearly reveals a strong flare decrease occurring during the entire observation with the initial rise happening before the EPIC/pn observation starts, showing a gradual decay characteristic of coronal flares in magnetically active stars. Flare events are commonly observed in young, magnetically active stars and are consistent with the elevated levels of coronal activity expected for members of nearby young moving groups.

\subsubsection{X-ray spectra}\label{subsect:XMMspectra}

To model the X-ray spectra, we use Cstat with the Bayesian X-ray Analysis package \citep[\texttt{BXA};][]{buchner2014}, which interfaces the nested sampling algorithm \texttt{UltraNest} \citep{buchner2021} with the \texttt{CIAO/Sherpa} fitting environment \citep{fruscione2006}. This Bayesian framework allows for robust exploration of parameter space and reliable uncertainty estimation, using prior definitions, which is particularly advantageous for spectra with limited photon statistics or strong parameter degeneracies. To constrain the properties of the X-ray emitting plasma, we fitted the EPIC/pn spectra using a two-component thermal {\texttt APEC} model, only for the one source with more than 350 net source counts (VCA~48; see Fig.~\ref{fig:1}), and a single-component {\texttt APEC} model for the remaining seven sources (see Appendix~\ref{appen:xmmSpectraAll}). Each APEC model consists of three parameters, namely the plasma temperature (kT), the global elemental abundance (Z), and the emission measure (EM). The temperature and emission measure are left free to vary, while the global abundance is fixed at 0.3 Z$_{\odot}$, the typical value for late-type stars \citep{favata2000,robrade2005,maggio2007}. 

To produce a Cstat fit with \texttt{BXA} we first defined priors for the free parameters, with an initial uniform prior for the temperature between 0.2-10\,keV, and a logarithmic uniform prior for the normalization between $1\times10^{-6}$ and $5\times10^{-3}$. The priors were adjusted individually depending on the quality of the fit after the initial selection. We report in Table~\ref{table:cstatResults} the values obtained for the best-fitting model of the eight spectra. 

\begin{table*}
\centering
    \begin{threeparttable}
    \caption{Best-fit parameters for the {\it XMM-Newton} EPIC/pn spectra, with values corresponding to upper and lower 90\% confidence ranges.}             
    \label{table:cstatResults}  
    \begin{tabular}{c ccc cc cc cc}
    \hline              
    \noalign{\smallskip}
        VCA & Detection & Counts & Count rate & log($L_{\rm x}$)$^a$ & log($L_{\rm x}/L_{\rm bol}$)$^a$ & kT$_1$ & EM$_1$ & kT$_2$ & EM$_2$\\
        ID & likelihood & & $\times$10$^{-3}$[cts s$^{-1}$] & [erg s$^{-1}$] & & [keV] & $\times$10$^{51}$[cm$^{-3}$] & [keV] & $\times$10$^{51}$[cm$^{-3}$] \\
        \noalign{\smallskip}
        \hline              
        \noalign{\smallskip}
        22 & 241 & 170$\pm$15 & 18.10$\pm$1.65 & 28.3 & -2.88 & 0.93$_{-0.13}^{+0.16}$ & 1.56$_{-0.27}^{+0.32}$\\
        \noalign{\smallskip}
        26 & 45.3 & 61$\pm$11 & 3.87$\pm$0.68 & 27.7 & -3.26 & 0.50$_{-0.15}^{+0.26}$ & 0.34$_{-0.12}^{+0.15}$&&\\
        \noalign{\smallskip}
        35 & 36.3 & 86$\pm$14 & 3.92$\pm$0.65 & 27.8 & -2.91 & 1.93$_{-0.51}^{+0.65}$ & 0.48$_{-0.14}^{+0.15}$&&\\
        \noalign{\smallskip}
        38 & 32.1 & 36$\pm$7 & 7.73$\pm$1.63 & 28.1 & -3.17 & 0.55$_{-0.20}^{+0.26}$ & 0.97$_{-0.29}^{+0.40}$&&\\
        \noalign{\smallskip}
        44 & 85.9 & 75$\pm$11 & 8.10$\pm$1.19 & 28.3 & -2.94 & 0.35$_{-0.08}^{+1.21}$ & 1.56$_{-0.35}^{+0.36}$&&\\
        \noalign{\smallskip}
        48$^b$ & 3044 & 1085$\pm$36 & 113$\pm$3 & 29.2 & -2.64 & 0.35$_{-0.03}^{+0.03}$ & 7.89$_{-0.89}^{+0.79}$ & 1.37$_{-0.08}^{+0.10}$ & 9.05$_{-0.83}^{+0.91}$  \\
        \noalign{\smallskip}
        53 & 28.0 & 38$\pm$8 & 7.81$\pm$1.71 & 28.2 & -3.80 & 3.18$_{-1.01}^{+1.22}$ & 1.66$_{-0.37}^{+0.43}$&&\\
        \noalign{\smallskip}
        57$^b$ & 219.6 & 145$\pm$14 & 30.38$\pm$2.95 & 28.7 & -2.80 & 1.61$_{-0.33}^{+0.87}$ & 5.71$_{-0.76}^{+0.72}$&&\\
        \noalign{\smallskip}
        \hline 
    \end{tabular}
    \begin{tablenotes}
        \small
        \item $^a$ Values of $L_{\rm x}$ calculated for the ROSAT band (0.1-2.4keV).
        \item $^b$ Flare was observed in the {\it XMM-Newton} broadband (0.2-12\,keV) lightcurve. 
    \end{tablenotes}
    \end{threeparttable}
\end{table*}

In order to facilitate comparison across different instruments and with previous studies, a posterior distribution of the X-ray flux was produced using \texttt{BXA} in the ROSAT band (0.1–2.4 keV), from which we obtain the mean X-ray flux and its 1$\sigma$ uncertainty. These fluxes were then combined with distances to derive the corresponding X-ray luminosities.

\subsection{ROSAT detections}\label{subsect:rosat}

To complement the X-ray data for the VCA from {\it XMM-Newton} and eROSITA, we cross-matched the proper motion corrected {\it Gaia} coordinates of the VCA catalog with the Second ROSAT all-sky survey (2RXS) source catalog. Using a search radius of 30" we identified ROSAT counterparts for 12 VCA members and candidates, however, none of these correspond to a star not detected with eROSITA or {\it XMM-Newton}. 

\subsection{Upper limits}\label{subsect:upperLimits}

Of the six stars without X-ray detections we derive upper limits for two of them, VCA~56 and VCA~50, using {\it XMM-Newton} observations. 
The upper limits for the three VCA candidates which had their {\it XMM-Newton} observation  affected by "Full Scientific buffer", as well as the candidate not targeted by {\it XMM-Newton} are obtained from eROSITA data. The upper limit count rates, X-ray luminosities, and fractional X-ray luminosities are reported in Table~\ref{tab:upperLimits}.

To obtain the EPIC/pn upper limits we defined circular regions in the sensitivity map created during source detection with radii of 30" centered on the star's coordinates, after having corrected them to the epoch of the {respective {\it XMM-Newton} observation. The mean count rate within each region was adopted as an upper limit to the source count rate. To convert these upper limit count rates into fluxes in the ROSAT band, we derived a $CF$ from the sources detected by {\it XMM-Newton}. Specifically, we used count rates from the source detection procedure and ROSAT-band fluxes from the spectral analysis of the eight stars detected with EPIC/pn to compute individual $CF$s, adopting their average value of 1.73$^{+0.47}_{-0.30}\times 10^{-12}$\,erg\,cm$^{-2}$\,cts$^{-1}$. This approach assumes similar spectral properties among the sources, which is reasonable given that the sample is dominated by late-type M stars (M4–M7) with only one earlier-type A9 star.

For the eRASS upper limits we follow the methodology described above. We extract count rates from sensitivity maps generated for the eROSITA sky tiles containing the VCA objects, using the \texttt{ESENSMAP} task from the eROSITA Science Analysis Software System (eSASS; version 240410.0.4). The conversion factor adopted in Sect.~\ref{subsection:eROSITA} for M stars is then applied to these count rates to derive X-ray fluxes in the ROSAT band. 

\begin{table}
    \centering
    \begin{threeparttable}
    \caption{Upper limits from {\it XMM-Newton} or eROSITA.}
    \label{tab:upperLimits}
    \begin{tabular}{c ccc}
    \hline              
    \noalign{\smallskip}
         VCA & Count rate & log(L$_{\rm x}$)$^a$ & log(L$_{\rm x}$/L$_{\rm bol}$)$^a$ \\
        ID & $\times$10$^{-3}$[cts s$^{-1}$] & [erg s$^{-1}$] & \\
        \noalign{\smallskip}
        \hline              
        \noalign{\smallskip}
        34 & $<$13.27 & $<$28.1 & $<$-2.92\\
        37 & $<$4.98 & $<$27.8 & $<$-3.40\\
        41 & $<$3.96 & $<$27.7 & $<$-3.01\\
        50 & $<$4.70 & $<$27.8 & $<$-3.19\\
        56 & $<$11.83 & $<$28.2 & $<$-6.45\\
        58 & $<$8.79 & $<$27.9 & $<$-2.77\\
        \noalign{\smallskip}
        \hline
    \end{tabular}
    \begin{tablenotes}
        \small
        \item $^a$ $L_{\rm x}$ is calculated for the ROSAT band (0.1-2.4keV).
    \end{tablenotes}
    \end{threeparttable}
\end{table}

\section{Discussion}\label{sect:discussion}

The high ($\sim$90$\%$) completeness of our X-ray census of the VCA allows for a robust characterization of the distributions of X-ray luminosities and fractional X-ray luminosities across nearly the full stellar mass range of the association, while minimizing biases introduced by non-detections. For other nearby associations, typical X-ray detection fractions are much lower, e.g. $\sim$30-73$\%$ for the Hyades \citep{freund2020,khamitov2025} to $\sim$30-50$\%$ for the Pleiades \citep{stauffer1994,khamitov2024}, where the range refers to different SpT groups. Our X-ray census of the VCA, therefore, is one of the most complete available to date for nearby young stellar associations.

\subsection{X-ray brightness across the VCA population}\label{subsect:XrayDiscussion}

\begin{figure}
    \centering
    \includegraphics[width=0.47\textwidth,trim=0.4cm 0.2cm 1.2cm 0.7cm,clip]{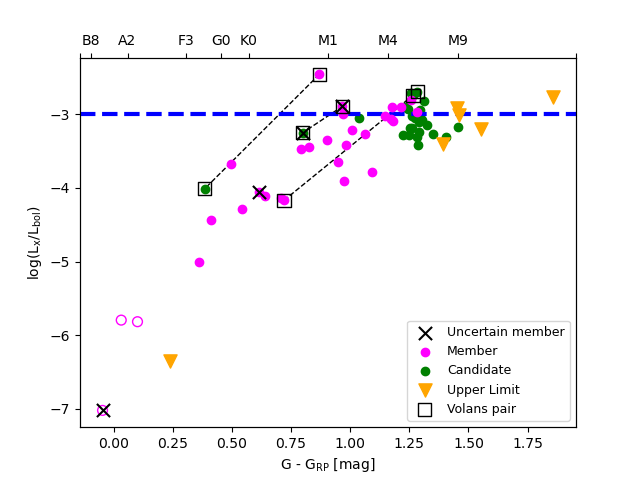}\hfill
    \includegraphics[width=0.47\textwidth,trim=0.4cm 0.2cm 1.2cm 0.7cm,clip]{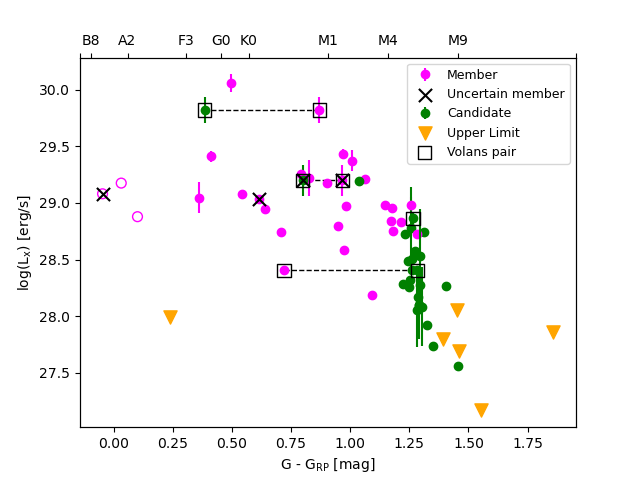}
    \caption{Fractional X-ray luminosity with the canonical saturation limit as the blue dashed line (\textit{upper panel}) and X-ray luminosity (\textit{lower panel}) of the VCA. Upper limits are indicated by yellow triangles. Unfilled markers indicate early type stars with an X-ray emission likely from an undetected companion. The vertical bars in the lower panel show the range of X-ray luminosities derived from observations with different instruments. 
    }
\label{fig:Tdiff}
\end{figure}

The X-ray luminosities of HPMs and CMs from the VCA are shown in Fig.~\ref{fig:Tdiff}, where vertical bars indicate the range of luminosities derived from observations with different instruments. Most VCA objects having a detection from a single instrument, while 18 objects were detected by two instruments (see Table~\ref{table:vcaCatalog}). 

The tendency for the lowest-luminosity objects to exhibit higher luminosities in the {\it XMM-Newton} observations than in eROSITA (Fig.~\ref{fig:LxComparison}) may arise from the different observing strategies of the two missions. The long {\it XMM-Newton} exposures have a greater likelihood of including recurrent flaring activity in addition to quiescent emission, whereas the shorter eROSITA observations sample a more limited portion of the variability cycle. This effect likely contributes to higher time-averaged luminosities in {\it XMM-Newton} for some sources.

We inspected the individual eROSITA light curves (see Appendix~\ref{appen:eROSITAlc}) of all VCA members and candidates with eROSITA counterparts. Using the standard deviation of the count rates as a variability metric, most VCA stars exhibit only modest variability ($\sigma \sim$ 0.01-0.08), whereas three sources stand out with substantially larger scatter: VCA~4 ($\sigma=0.19$) and the optically resolved CPM binary composed of VCA~28 and VCA~11 ($\sigma=0.21$). These are also the three most X-ray luminous systems in the association (log($L_{\rm x}) > 29.5$ erg\,s$^{-1}$ in Fig.~\ref{fig:Tdiff}), suggesting that enhanced short-term variability might be a factor contributing to their elevated time-averaged X-ray emission.

The brightest X-ray source is VCA~4, a G0-type HPM, which exhibits elevated X-ray luminosities in both the eROSITA and ROSAT observations. There is no current evidence for resolved or unresolved multiplicity that could account for its enhanced emission. Its eROSITA light curve (Appendix~\ref{appen:eROSITAlc}) reveals repeated flaring on the few-hour cadence sampled by eRASS, indicating that short-term variability contributes to the observed luminosity. However, even after accounting for the estimated flare contribution, the inferred quiescent luminosity remains slightly higher than that of most association members (log($L_{\rm x,qui})\,=\,29.8$ [erg/s]). Together with the consistently elevated luminosities measured over the $\sim$30 year baseline between ROSAT and eROSITA, this suggests that VCA~4 is intrinsically the most X-ray-active star in the association, although the origin of its enhanced activity remains unclear.

The second most luminous stars, VCA~28 (a F5V-type HPM) and VCA~11 (a K7-type CM), form an optically resolved CPM binary that shares a single X-ray counterpart. Although the observed luminosity was divided equally between the two stars to approximate their individual contributions (see Sect.~\ref{subsection:eROSITA}), both components remain more X-ray luminous compared with the rest of the association. Given their wide separation, binary interaction is not expected to enhance magnetic activity. Their elevated X-ray luminosities may therefore reflect the upper end of the intrinsic activity distribution within the association, consistent with the inferred quiescent luminosity being slightly higher than that of most association members ($\log(L_{\rm x,qui}) = 29.6$ erg,s$^{-1}$), similarly to VCA~4.

\begin{figure}
    \centering
    \includegraphics[width=0.95\columnwidth,trim=0.3cm 0.1cm 1.3cm 1.5cm,clip]{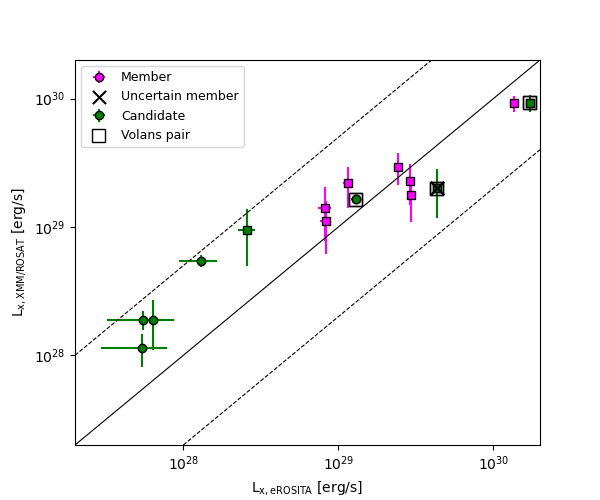}
    \caption{Comparison between X-ray luminosity values obtained for the 18 VCA objects with detection from different instruments. All objects have an eROSITA detection with an additional detection from either ROSAT (squares) or {\it XMM-Newton} (circles). Variation of half an order of magnitude is illustrated with the dotted black lines.
    }
    \label{fig:LxComparison}
\end{figure}

\begin{figure*}
    \centering
    \includegraphics[width=\textwidth,trim=0.2cm 0.2cm 0.2cm 0.2cm,clip]{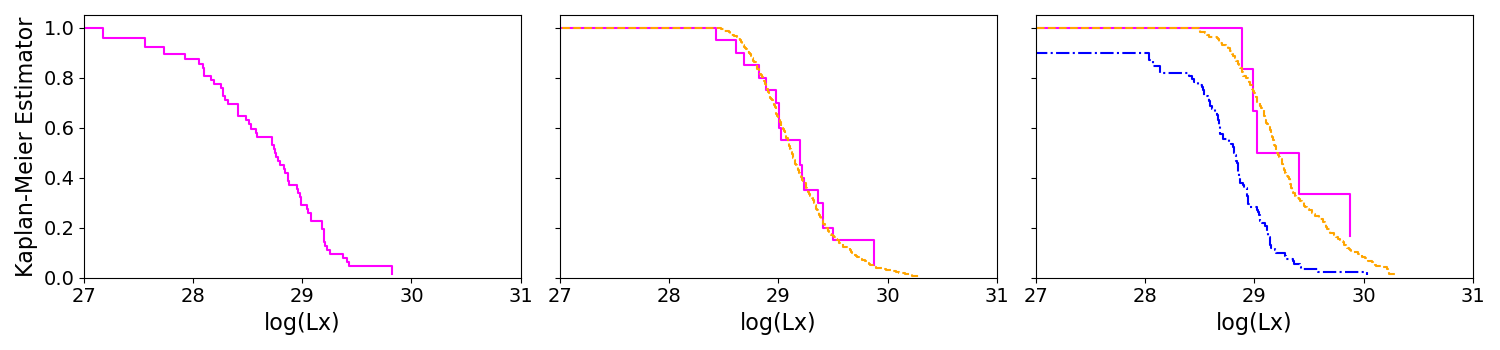}
    \caption{Kaplan-Meier XLF of the VCA (solid magenta). {\it Left panel:} For the entire association. {\it Middle panel:} For stars with spectral types F5 to M3, with the XLF of the Pleiades \citep[dashed yellow line;][]{khamitov2024} in the same spectral type range for comparison. {\it Right panel:} For stars with spectral types F5 to K4, with the XLF of the Pleiades and Hyades \citep[dotted blue line;][]{freund2020} in the same spectral type range for comparison.}
    \label{fig:XLF}
\end{figure*}

The five undetected late-type CMs have upper limits consistent with the expected $L_{\rm x}$ for their spectral type. A similar value for the upper limit to $L_{\rm x}$ was measured for VCA~56, an A9-type HPM, which remained  undetected in all available X-ray observations. A non-detection is, indeed, expected for a likely single late-A star, since stars of this type lack the outer convective envelopes required to sustain strong solar-like dynamos. As a result, intrinsic coronal X-ray emission from A-type stars is generally weak or absent, with X-ray detections reported for late-A stars consistent with the derived upper limit (log($L_{\rm x}$)\,$\leq$\,28.1 [erg/s]; \citealt{gunther2022}). In this sense, the more unusual systems are the early-type HPMs that are detected in X-rays: VCA~39, VCA~23, and VCA~1 (see unfilled markers in Fig.~\ref{fig:Tdiff}). Recent multiplicity analysis by \cite{gratton2024} identified all of these three as candidate binaries which remain unresolved with {\it Gaia}. Two of them were flagged due to an elevated {\it Gaia} RUWE value, and one through a large RV amplitude. These suggested companions have dynamical masses consistent with G or K-type stars, with the observed X-ray luminosities being typical for these spectral types at the age of the VCA. The detected X-ray emission in these early-type systems, therefore, most likely originates from their lower-mass companions.

We use the stellar masses derived from the best-fit BHAC isochrone in Sect.~\ref{subsect:isochrones} to define three spectral type ranges relevant for the study of low-mass stars in young associations \citep{richey2023}: K stars (0.6-0.9 M$_{\odot}$), early-M stars (M1 to M3, 0.35-0.6 M$_{\odot}$), and late-M stars (M4 to M9, 0.08-0.35 M$_{\odot}$).

\begin{table}
    \centering
    \caption{Median values for the X-ray luminosity and fractional X-ray luminosity of K and M stars in the VCA.}
    \label{tab:medianvalues}
    \begin{tabular}{c ccc}
    \hline              
    \noalign{\smallskip}
          & K stars & early-M stars & late-M stars\\
        \noalign{\smallskip}
        \hline              
        \noalign{\smallskip}
        $L_{\rm x}$ [erg/s] & 29.1 & 28.9 & 28.4\\
        log($L_{\rm x}/L_{\rm bol}$) & -3.68 & -3.30 & -3.10\\
        \noalign{\smallskip}
        \hline
    \end{tabular}
\end{table}

The median X-ray luminosity decreases from K-, over early-M to late-M stars (see Table~\ref{tab:medianvalues}). This decline and the values we measure for the spectral subclasses are consistent with the prediction of spin evolution models for the age of the VCA \citep{johnstone2021} and, hence, go back to the combination of mass and age dependence of the stellar rotation rate. While the X-ray luminosity decreases for later spectral types the $L_{\rm x}/L_{\rm bol}$ ratio increases (see Table~\ref{tab:medianvalues}). The canonical saturation limit log($L_{\rm x}/L_{\rm bol}$) $\approx$ -3 \citep[e.g.][]{pizzolato2003,zuckerman2004,wright2011} is reached only for the M stars. 

\subsection{X-ray luminosity function}\label{subsect:XlfDiscussion}

To compare the VCA with other stellar populations, we constructed X-ray luminosity functions (XLFs). We used the ASURV package (version 1.2; \citealt{feigelson1985,isobe1986,lavalley1992}), which is well suited for the study of censored data sets containing upper limits. The XLF includes luminosity values for 57 HPMs and CMs from the VCA, together with upper limits for one HPM and five CMs, is displayed in Fig.~\ref{fig:XLF} (left panel). For the four optically resolved CPM binaries, we assumed that the X-ray luminosity is divided equally among the components (see Sect.~\ref{subsection:eROSITA}). 

Although the X-ray census of the VCA is nearly complete down to very low masses, this is not yet the case for other stellar populations. We, therefore, restrict the XLF comparison to spectral type intervals in which the X-ray census of other clusters and associations are highly complete. 

First, we use the Pleiades cluster given its similar age and recently improved X-ray census. With the eROSITA survey of the cluster \citep{khamitov2024}, which is $\sim$99$\%$ located in the half-sky defined by Russian eROSITA data rights, completeness in the F5 to M3 range has reached $\sim$93$\%$. This value decreases due to the eROSITA sensitivity limit, to $\sim$33$\%$ for earlier types and $<$\,56$\%$ for later types. Restricting both samples to the F5-M3 range (Fig.~\ref{fig:XLF}, middle panel), where the VCA is 100$\%$ complete, we find that the XLFs of the VCA and Pleiades are indistinguishable. Two-sample survival-analysis tests for censored data, as implemented in ASURV, yield probabilities of $p$\,=\,0.76 (Gehan’s generalized Wilcoxon test) and $p$\,=\,0.74 (logrank and Peto–Peto tests), indicating, indeed, no statistically significant difference between the two populations. 

We also compared the XLF of the VCA to that of the Hyades. To construct the XLF of the Hyades we use the X-ray catalog of \cite{freund2020}, where data from ROSAT, {\it Chandra}, and {\it XMM-Newton} is combined. The catalog achieves a $\sim$85$\%$ completeness over the F5 to K4 range, comparable to a 100$\%$ and 95$\%$ completeness obtained in this range for the VCA and Pleiades respectively. In contrast to the Pleiades comparison, the VCA and Hyades XLFs (Fig.~\ref{fig:XLF}, right panel) differ significantly according to all applied two-sample tests ($p\,<$\,2.4$\times$10$^{-3}$). These results quantitatively confirm that F5 to K4 stars in the VCA and the Pleiades are substantially more X-ray active than the older Hyades stars.

\subsection{Updated age-decay of X-ray activity}\label{subsect:richeyDiscussion}

To place the VCA in the broader evolutionary sequence, we compare our results with the analysis of \cite{richey2023}, who examined X-ray fluxes scaled to a distance of 10\,pc for K, early-M, and late-M stars in nine NYMGs (TW Hydra, 10$\pm$3\,Myr; Carina, 13.3$\pm$1\,Myr; $\beta$ Pictoris, 24$\pm$3\,Myr; Columba, 42$\pm$3\,Myr; Tuc-Hor, 45$\pm$4\,Myr; AB Dor, 149$_{-19}^{+51}$\,Myr; Ursa Majoris, 414$\pm$23\,Myr; Praesepe, 600\,Myr; Hyades, 650$\pm$70 Myr) and the field (5000\,Myr). However, these samples were affected by highly incomplete memberships with X-ray censuses dominated by upper limits. The VCA, therefore, provides an important new anchor point with substantially improved completeness.

\begin{table}
    \centering
    \caption{Interquartiles of the VCA flux distribution at 10pc in erg\,s$^{-1}$\,cm$^{-2}$.}
    \label{tab:interquartiles}
    \begin{tabular}{c ccc}
    \hline              
    \noalign{\smallskip}
          & K stars & early-M stars & late-M stars\\
        \noalign{\smallskip}
        \hline              
        \noalign{\smallskip}
        25th & 2.5$\times$10$^{-11}$ & 2.0$\times$10$^{-11}$ & 4.7$\times$10$^{-12}$\\
        50th & 9.6$\times$10$^{-12}$ & 1.3$\times$10$^{-11}$ & 2.6$\times$10$^{-12}$\\
        75th & 4.6$\times$10$^{-12}$ & 5.2$\times$10$^{-12}$ & 1.2$\times$10$^{-12}$\\
        \noalign{\smallskip}
        \hline
    \end{tabular}
\end{table}

\begin{figure*}
    \centering
    \includegraphics[width=0.325\textwidth,trim=6.2cm 0.2cm 5.2cm 0.2cm,clip]{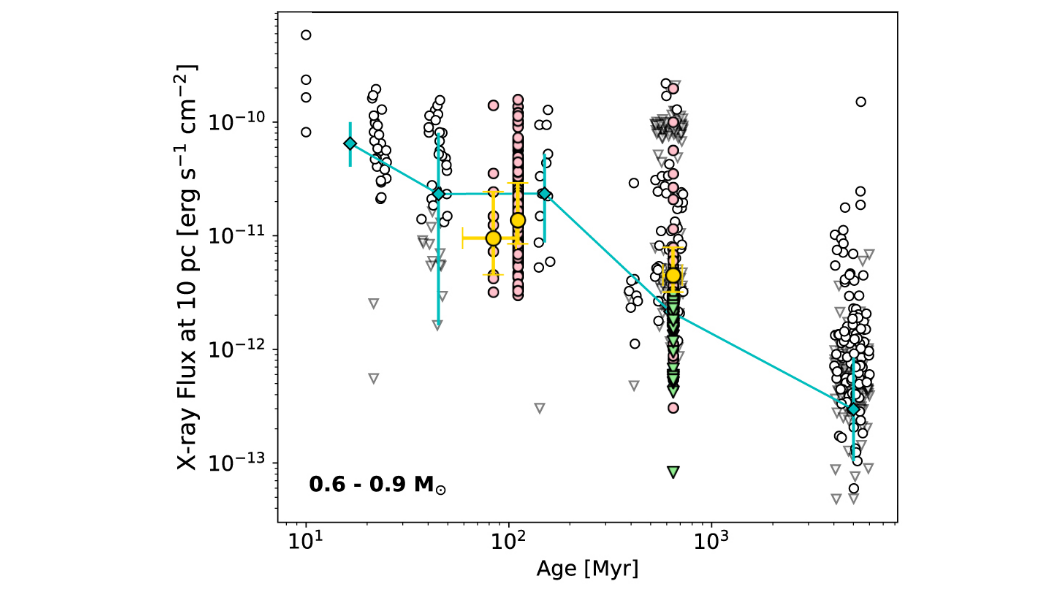}
    \includegraphics[width=0.33\textwidth,trim=5.5cm 0cm 5.8cm 0.8cm,clip]{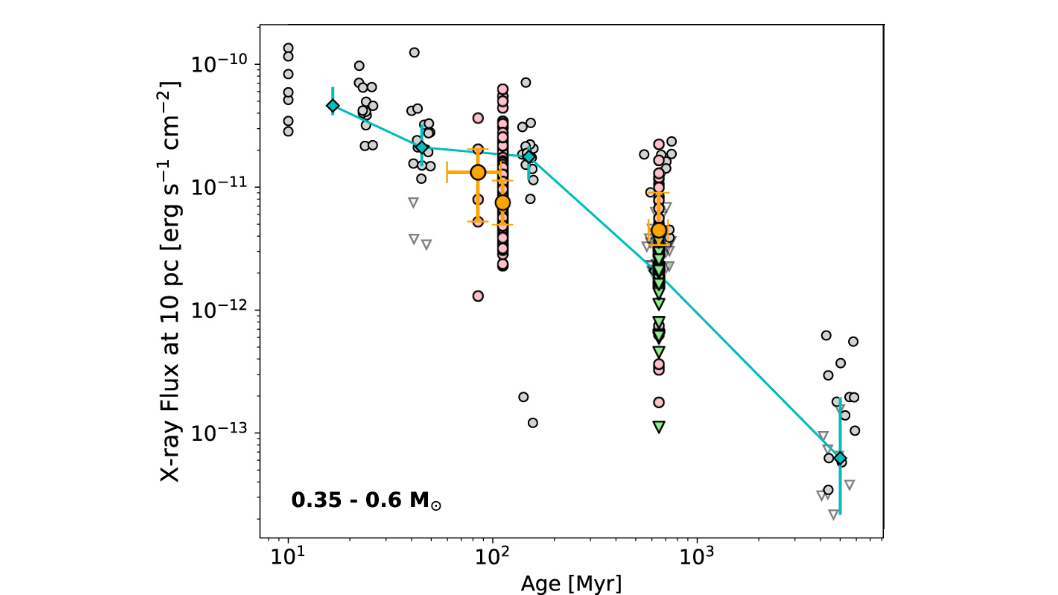}
    \includegraphics[width=0.33\textwidth,trim=5.6cm 0cm 5.6cm 0.7cm,clip]{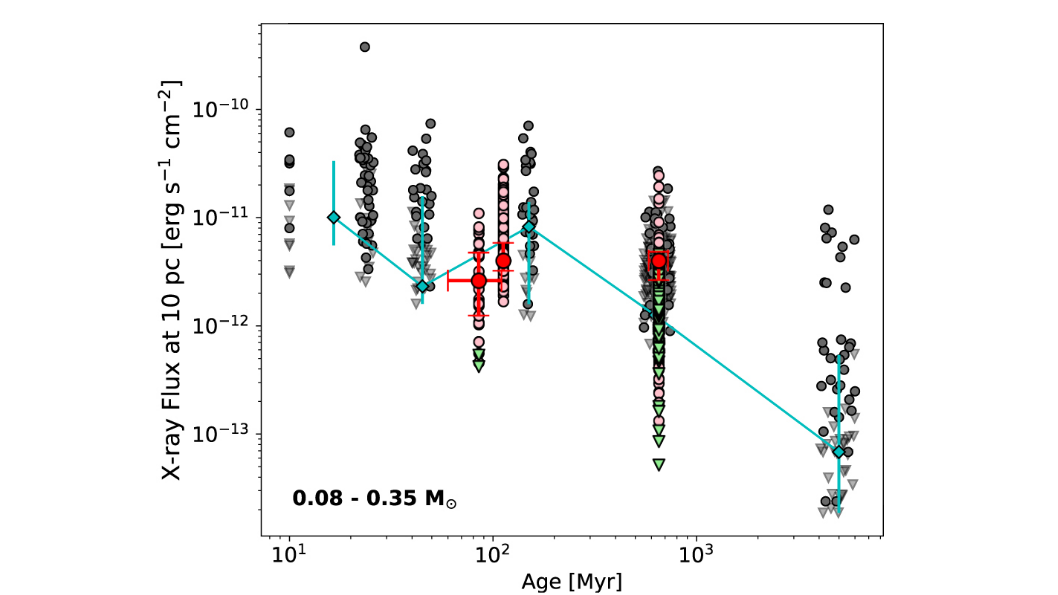}
    \caption{Age-decay of X-ray emission for K stars (left), early-M stars (middle), and late-M stars (right), adapted from \cite{richey2023}. Newly overlaid data are highlighted with filled symbols, where circles indicate detections and triangles denote upper limits. The original relations and distributions from their work are reproduced here, with the VCA, Pleiades \citep{khamitov2024}, and Hyades \citep{freund2020} overlaid to evaluate the impact of updated and more complete X-ray censuses on the empirical age-decay relation. 
    }\label{fig:richey1}
\end{figure*}

\begin{figure}
    \centering
    \includegraphics[width=0.85\columnwidth,trim=1.2cm 0cm 1.9cm 0cm,clip]{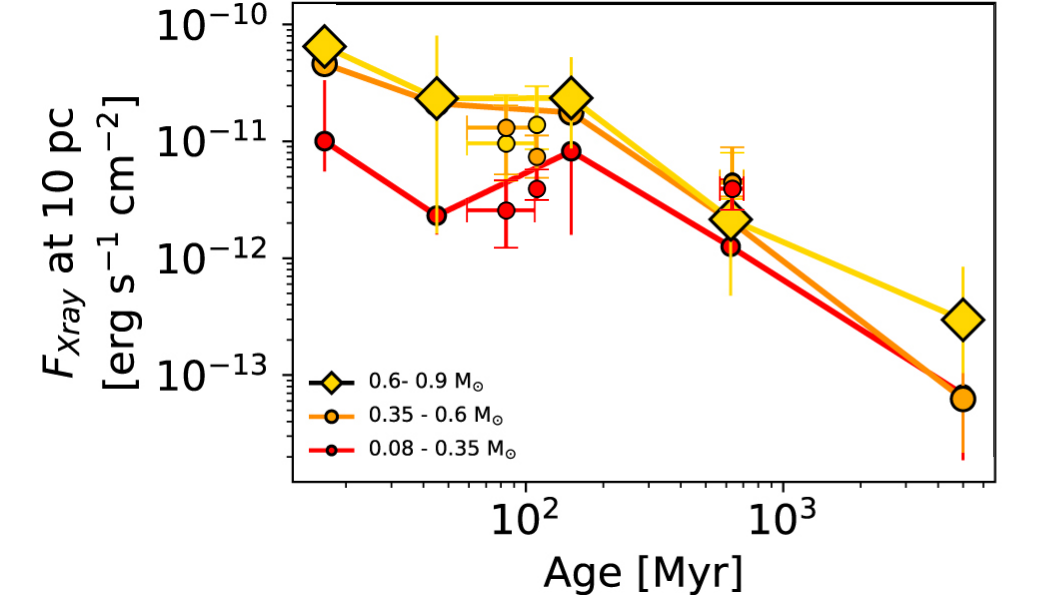}
    \caption{Combination of the three spectral-type relations from Fig.~\ref{fig:richey1} into a single figure, representing the median value with error bars showing the percentiles.  
    }\label{fig:richey2}
\end{figure}

The interquartile ranges (25th to 75th) of the VCA flux distribution at 10\,pc are shown in Table~\ref{tab:interquartiles}, corresponding to an intrinsic spread of approximately 0.6-0.7 dex within each spectral bin. As in other NYMGs, this substantial dispersion likely reflects a combination of short-term variability, rotation, and binarity. Notably, the most X-ray luminous K and early-M stars in the VCA are part of resolved CPM binaries, while the brightest late-M VCA star was observed in a flaring state with {\it XMM-Newton}. We retain these systems in order to maintain consistency with \cite{richey2023}, where analogous effects were not explicitly removed.

In Fig.~\ref{fig:richey1}, we place the VCA on the original relations of \cite{richey2023}, using the 75th, 50th, and 25th percentiles derived with ASURV (see Table~\ref{tab:interquartiles}), as well as the individual stars. We proceed analogously with the updated X-ray census of the Pleiades \citep{khamitov2024} and Hyades \citep{freund2020}. The results indicate a likeness between the X-ray brightness distribution of the VCA and the similar-aged Pleiades, which however breaks for the late-M stars. This is probably a signature of the incompleteness of the eRASS sample for the Pleiades, reinforcing the value of our new X-ray census of the VCA as an important calibration point at $\sim 100$\,Myr. Comparing the VCA (and the Pleiades) to its immediate neighbors in terms of age (Columba and Tuc-Hor with $\sim 40$\,Myr and AB\,Dor with $\sim 150$\,Myr), the normalized X-ray flux of the VCA appears to be lower than both age bins. Within the percentile uncertainties, the VCA is still consistent with the X-ray evolution identified by \cite{richey2023}. However, the uncertainties reported by \cite{richey2023} (light blue vertical lines in Fig.~\ref{fig:richey1}) are very large for some of their samples. Especially, AB\,Dor (in particular for the late-M stars) is counteracting the expected smooth age decay. This might be attributed to the incompleteness not only of the X-ray catalog of this association but also a consequence of the fact that the number of known M-stars members in AB\,Dor is smaller than expected from the initial mass function \citep{gagne2018c}. As far as the Hyades are concerned, in the census by \cite{freund2020} the upper limits have been decreased by up to one order of magnitude but the average X-ray brightness inferred from their flux distribution is slightly higher than that presented by \cite{richey2023}. A further examination of the discrepancies between the two Hyades studies is not in the scope of our work. In any case, both Hyades studies are highly dominated by upper limits.

In Fig.~\ref{fig:richey2} we summarize the updated X-ray flux evolution for the three SpT ranges by integrating the original analogous figure from \cite{richey2023} with the new data for VCA, the Pleiades and the Hyades. From Fig.~\ref{fig:richey2} it is evident that the X-ray activity level at ages of $100 \pm 50$\,Myr still remains ambiguous. The current membership list and X-ray census of AB Dor appears insufficiently complete for defining the empirical age-activity relation with the same level of confidence as the VCA or the Pleiades, which present more reliable constraints.

\section{Conclusions}\label{sect:conclusions}

We have presented an updated membership census of the VCA and the first nearly-complete X-ray survey of a NYMG, combining {\it Gaia} DR3 astrometry, as well as supplementary archival RV measurements, with data from eROSITA, ROSAT, and a dedicated {\it XMM-Newton} programme.

Our revised census identifies 29 HPMs, 30 CMs, three uncertain members, and one uncertain candidate, while rejecting three previous candidates as being part of the VCA. Isochronal analysis of the association supports an age range of $80\pm20$\,Myr, consistent with previous determinations. This confirms the VCA as a population slightly younger than, or comparable in age to, the Pleiades.

We detect X-ray emission from 28 HPMs, 25 CMs, three uncertain members, and one uncertain candidate and derive upper limits for the remaining HPM and five CMs, yielding an overall X-ray completeness of $\sim$90$\%$. This places the VCA as the most comprehensively characterized NYMG in X-rays. The low-mass stars in the VCA display fractional X-ray luminosities close to the canonical saturation level.

As a central result of this work, we use the VCA to improve empirical constraints on the evolution of X-ray activity with age in NYMGs. The VCA represents a particularly valuable association in this context, as it presents a highly complete membership census, down to 0.2,M$_{\odot}$, and a correspondingly high level of X-ray completeness. This makes it one of the most robust benchmarks currently available for calibrating the age-activity relation at young ages, where many previous samples have been limited by incomplete memberships and X-ray censuses dominated by upper limits.

Comparisons of XLFs from recently published works focused on the Pleiades and Hyades show, for different spectral type ranges, that the VCA is statistically consistent with the Pleiades. Whether the X-ray activity remains constant or slightly declines from $\sim10-100$\,Myrs now mainly rests on a better definition of the X-ray census for the intermediate stage at $\sim 50$\,Myrs. The new Hyades X-ray census is consistent with the previous one except for the late-M stars where the average flux appears brighter than inferred before. Even this updated Hyades X-ray study is highly incomplete for the full M spectral type range. As a consequence, also for ages of  $\sim 500$\,Myr the X-ray luminosity distribution continues to be poorly constrained. 

Taken together, these results establish the VCA as a key system for calibrating the early evolution of X-ray activity. As demonstrated in our comparison with literature samples, the VCA significantly strengthens empirical constraints on the age decay of X-ray activity relation in the critical 50–150\,Myr regime, bridging the gap between very young star-forming regions and older open clusters. The association therefore plays a central role in nailing down the onset and timescale of coronal activity decay, and it provides an essential reference point for future studies of activity evolution, rotation-activity relations, and the high-energy environments of young planetary systems.
However, clear conclusions on the age-decay can be obtained only if a similarly high completeness level is reached for associations representing the other age groups.

\begin{acknowledgements}
Daniela Muñoz-Giraldo acknowledges financial support by the Bundesministerium für Wirtschaft und Energie through the Deutsches Zentrum für Luft- und Raumfahrt e.V. (DLR) under grant number FKZ 50 OR 2501. This work is based on observations obtained with {\it XMM-Newton}, an ESA science mission with instruments and contributions directly funded by ESA Member States and NASA. This work is based on data from eROSITA, the soft X-ray instrument aboard SRG, a joint Russian-German science mission supported by the Russian Space Agency (Roskosmos), in the interests of the Russian Academy of Sciences represented by its Space Research Institute (IKI), and the Deutsches Zentrum f\"{u}r Luft- und Raumfahrt (DLR). The SRG spacecraft was built by Lavochkin Association (NPOL) and its subcontractors, and is operated by NPOL with support from the Max Planck Institute for Extraterrestrial Physics (MPE). The development and construction of the eROSITA X-ray instrument was led by MPE, with contributions from the Dr. Karl Remeis Observatory Bamberg \& ECAP (FAU Erlangen-Nuernberg), the University of Hamburg Observatory, the Leibniz Institute for Astrophysics Potsdam (AIP), and the Institute for Astronomy and Astrophysics of the University of T\"{u}bingen, with the support of DLR and the Max Planck Society. The Argelander Institute for Astronomy of the University of Bonn and the Ludwig Maximilians Universit\"{a}t Munich also participated in the science preparation for eROSITA. The eROSITA data shown here were processed using the eSASS/NRTA software system developed by the German eROSITA consortium. This work has made use of data from the European Space Agency (ESA) mission {\it Gaia} (\url{https://www.cosmos.esa.int/gaia}), processed by the {\it Gaia} Data Processing and Analysis Consortium (DPAC, \url{https://www.cosmos.esa.int/web/gaia/dpac/consortium}). Funding for the DPAC has been provided by national institutions, in particular the institutions participating in the {\it Gaia} Multilateral Agreement. This publication makes use of VOSA, developed under the Spanish Virtual Observatory (\url{https://svo.cab.inta-csic.es}) project funded by MCIN/AEI/10.13039/501100011033/ through grant PID2020-112949GB-I00. VOSA has been partially updated by using funding from the European Union's Horizon 2020 Research and Innovation Programme, under Grant Agreement n$^\circ$ 776403 (EXOPLANETS-A).

\end{acknowledgements}

%
%

\bibliography{references.bib}
\bibliographystyle{aa} 

\begin{appendix}
\onecolumn
\section{Galactic positions and space velocities of the VCA}\label{appen:cornerPlots}

\begin{figure*}
    \centering
    \includegraphics[width=\textwidth,trim=0.2cm 0.2cm 0.2cm 0.2cm,clip]{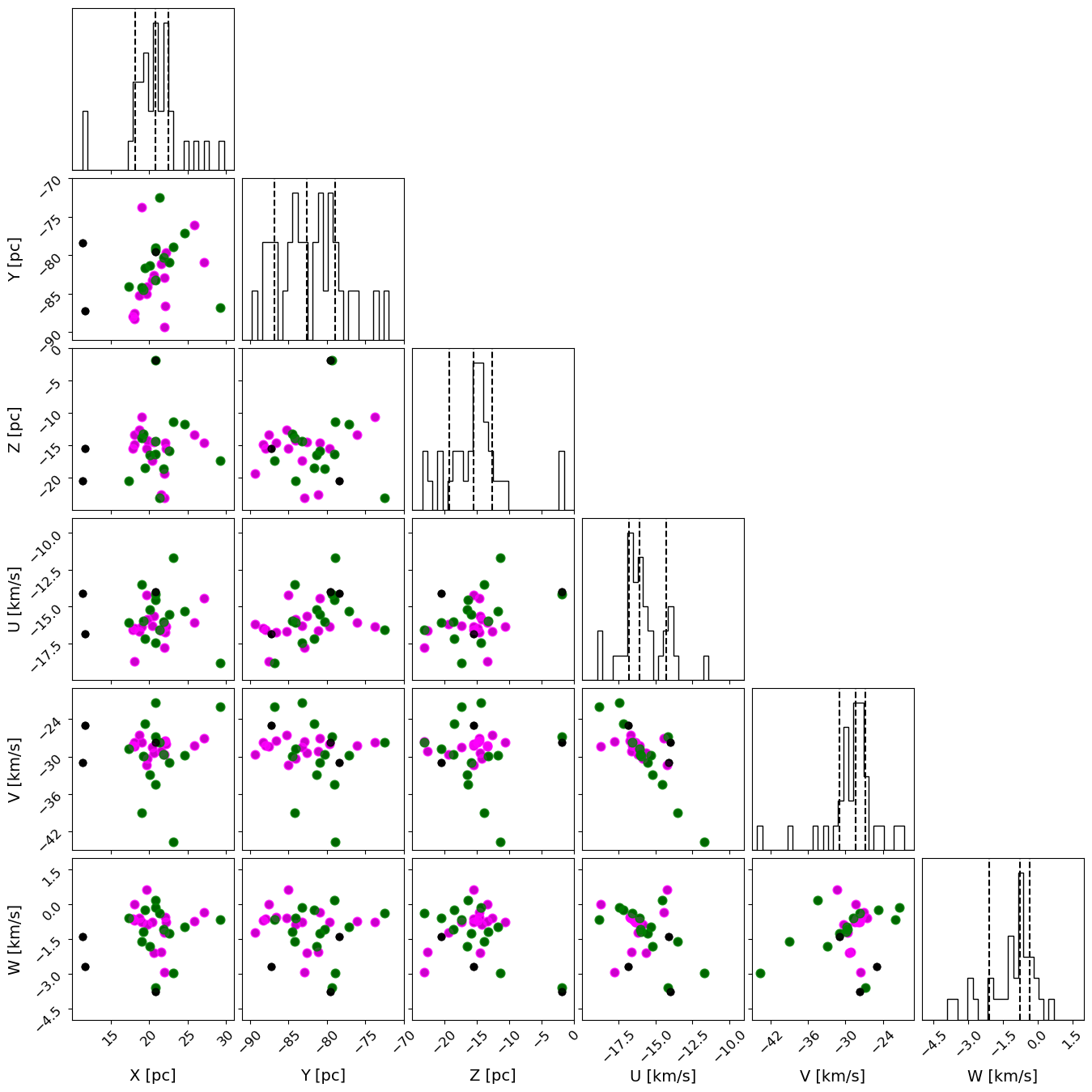}
    \caption{Galactic positions ({\it XYZ}) and space velocities ({\it UVW}) of the VCA members from \citetalias{gagne2018a} (in magenta), members from \citetalias{gagne2018a} that are not consistent with the VCA and therefore considered uncertain (in black) and previous candidates that we newly confirm as members (in green). Histograms show the median value and 1$\sigma$ boundaries.}
    \label{fig:vcaCorner}
\end{figure*}

Fig.~\ref{fig:vcaCorner} presents a corner plot of the Galactic positions ({\it XYZ}) and space velocities ({\it UVW}) of the 13 new kinematically confirmed members of the VCA. The clustering visible in both position and velocity space demonstrates that the new members are consistent with the known locus of the VCA. There are some outliers apparent in the projected dimensions. All three uncertain members show up as outliers compared to the locus of the VCA at least in one dimension, both {\it Z}-position and {\it W}-velocity for VCA~15, and only {\it X}-position for VCA~1 and VCA~6, while two new members, VCA 48 and VCA 62, appear as outliers with respect to the {\it V}-velocity locus of the VCA.

The key difference between these two new confirmed members and the three uncertain members is the membership probability derived using BANYAN $\Sigma$, which are consistent with the VCA for the earlier and with the field for the latter. Because of this, we maintain the classification of these two systems as new kinematically confirmed members of the VCA.

\section{Catalog of the VCA}\label{appen:vcaCatalog}

Catalog of the VCA divided into HPMs, uncertain members, CMs and uncertain candidates (see Sect.~\ref{sect:census} for more details). For each object we report the: VCA ID used in \citetalias{gagne2018a}, \cite{bailer2021} distances, spectral type, bolometric luminosity and effective temperature derived in Sect.~\ref{sect:stellarParameter}, BANYAN $\Sigma$ membership probability (see Sect.~\ref{subsubsect:kin_test}), and X-ray detection.\\

\begin{longtable}{ll c cc ccc cc}
    \caption{Catalog of VCA members and candidates.}       
    \label{table:vcaCatalog}   \\
    \hline
    \noalign{\smallskip}
    VCA & Designation & $d$ & RV & Ref. & SpT & $L_{\rm bol}$\, & T$_{\rm eff}$ & $P_{mem}$ & X-ray \\
    ID & & [pc] & [km/s] & &  & [$L_{\odot}$] & [K] &  & detection\\
    \hline
    \noalign{\smallskip}
    \endhead
    \multicolumn{3}{l}{Highly probable Member} & & & & & & \\
    \hline 
    \noalign{\smallskip}
    23& HD 82406&83.0& 29.0$\pm$4.4& {\it d} & A0V &19.96&	7000& 0.994$^*$& eR\\
    39& HD 83523&87.0& 16.2$\pm$2.2& {\it e} & A2V &12.02	&7000& 0.985$^*$& eR\\
    56& HD 83948&85.4& 25.5$\pm$0.3& {\it a} & A9IV/V &5.74	&7000& 0.998$^*$&  ulX\\
    2& HD 80563&91.0& 24.1$\pm$0.2& {\it a,b} & F3V &2.95&	6500& 0.999$^*$& R, eR\\
    3& HD 83946&90.5& 22.2$\pm$0.5& {\it a} & F5V &1.86	&6200& 0.999$^*$& R, eR\\
    4& HD 309681&88.5& 26.8$\pm$0.2& {\it a,b} & G0 &1.40	&5700& 0.955$^*$& R, eR\\
    5& CD-67 852&81.4& 21.3$\pm$0.2& {\it a} & (K0) &0.61&	5500& 0.999$^*$& eR\\
    7& TYC 8953-1289-1&84.1& 22.3$\pm$0.3& {\it a} & (K4) & 0.29	&4900& 0.999$^*$& eR\\
    8& TYC 8950-1447-1&76.8& 22.7$\pm$0.3& {\it a,b} & (K6) &0.20	&4600& 0.999$^*$& eR\\
    9& TYC 9210-1818-1$^a$&86.5& 20.8$\pm$0.3& {\it a} & (K6) &0.20&	4600& 0.998$^*$& eR\\
    10& UCAC4 120-020514&87.5& 23.4$\pm$0.3& {\it a,b} & (K8) & 0.12	&4300& 0.999$^*$& R, eR\\
    55& UCAC4 092-015404&79.0& 21.1$\pm$0.6& {\it a} & (K8) &0.137	&4400& 0.990$^*$& eR\\
    11& UCAC4 125-022457$^b$&86.4& 24.7$\pm$0.7& {\it c} & K7 &0.10&	4100& 0.999$^*$& R, eR\\
    12& UCAC4 136-023859&88.2& 22.2$\pm$1.9& {\it a,b} & (M1) &0.09	&4000& 0.999$^*$& eR\\
    13& UCAC4 137-022243&90.4& 23.8$\pm$1.8& {\it a,b} & (M1) &0.10	&3700& 0.997$^*$& R, eR\\
    14& UCAC4 104-024769&88.9& 22.2$\pm$1.0& {\it a,b} & (M2) &0.08&	3800& 0.943$^*$& eR\\
    16& UCAC4 104-024773&86.9& 23.6$\pm$1.0& {\it a,b} & (M2) &0.07	&3800& 0.996$^*$& eR\\
    17& UCAC4 129-021088&87.5& 25.7$\pm$3.1& {\it a,b} & (M2) &0.07	&3800& 0.999$^*$& R, eR\\
    18& UCAC4 118-021664&93.8& 24.6$\pm$1.7& {\it a,b} & (M2) &0.08	&3600& 0.997$^*$& eR\\
    19& UCAC4 135-021824&91.1& 23.6$\pm$3.8& {\it a} & (M2) &0.06	&3800& 0.998$^*$& eR\\
    24& UCAC4 121-021394&85.1& 28.2$\pm$4.6& {\it a} & (M4) &0.019&	3300& 0.999$^*$& eR\\
    29& UCAC4 111-021741&85.0& 24.1$\pm$6.4 & {\it a}& (M4) &0.021	&3300& 0.999$^*$& eR\\
    31& UCAC4 118-020952&85.6& 19.6$\pm$4.6& {\it a} & (M4) &0.019	&3300& 0.999$^*$& eR\\
    32& UCAC4 107-026470&92.9& 14.7$\pm$4.9& {\it a} & (M4) &0.026	&3400& 0.943$^*$& eR\\
    53& UCAC4 133-023598&87.4& 25.6$\pm$2.1& {\it a} & (M4) &0.025	&3500& 0.999$^*$&  X\\
    61& UCAC4 121-018181&87.9& 24.3$\pm$3.2& {\it a} & (M4) &0.016	&3200& 0.998$^*$& R, eR\\
    62& UCAC4 125-026881&82.7& 38.7$\pm$4.7& {\it a} & (M4) &0.014	&3200& 0.932$^*$& eR\\
    20& 2MASS J10101695-6616006&81.5& 23.6$\pm$5.5& {\it a} & (M5) &0.013&	3100& 0.999$^*$& eR\\
    48& KPP 3788B$^c$&87.0& 34.9$\pm$6.5& {\it a,b} & (M5) &0.018&	2700& 0.996$^*$&  X, eR\\
    \hline 
    \noalign{\smallskip}
    \multicolumn{3}{l}{Uncertain member} && & & & \\
    \noalign{\smallskip}
    \hline 
    \noalign{\smallskip}
    1& c Car&89.7& 22.7$\pm$0.5 & {\it a} & B8II & 150.3	&7000 & 0.556$^*$&eR\\
    6& TYC 8933-327-1&81.8& 28.0$\pm$0.6& {\it c} & (K4) &0.32&	5100& 0.079$^*$& eR\\
    15& UCAC4 157-043712$^d$&82.3& 23.4$\pm$1.5& {\it a,b} & (M2) &0.07	&3800& 0.553$^*$& R, eR\\
    \hline 
    \noalign{\smallskip}
    \multicolumn{3}{l}{Candidate member} && & & & \\
    \noalign{\smallskip}
    \hline 
    \noalign{\smallskip}
    28& HD 83359$^b$&86.8& 22.9$\pm$1.1 && F5V &3.63	&6300& 0.999& R, eR\\
    27& UCAC4 105-027776&82.9& 21.9$\pm$1.1&& (M3) &0.046	&3600& 0.998& eR\\
    36& KPP 3788A$^c$&87.3& 23.5$\pm$1.1 && (M4) &0.021	&3500& 0.999& X, eR\\
    21& UCAC4 133-019108&78.1&23.7$\pm$1.1&& (M5) &0.010	&3200& 0.983& eR\\
    22& 2MASS J08485563-6113261&86.6& 24.9$\pm$1.2 && (M5) &0.004&	3100& 0.974& X, eR\\
    25& 2MASS J09244337-6856223&85.0& 22.5$\pm$1.1 && (M5) &0.007&	3200& 0.998& eR\\
    30& 2MASS J09090086-6826022&88.9& 22.6$\pm$1.1 && (M5) &0.010	&3100& 0.995& eR\\
    33& 2MASS J09312193-6419239&94.4& 24.1$\pm$1.1 && (M5) &0.006	&3100& 0.988& eR\\
    37& 2MASS J09223111-6206070&89.5& 24.3$\pm$1.2 && (M5) &0.004	&2900& 0.996& uleR\\
    38& 2MASS J09244301-6254468&89.1& 24.3$\pm$1.2 && (M5) &0.004	&3100& 0.997&  X, eR\\
    40& 2MASS J09181573-6310540&85.7& 23.9$\pm$1.1 && (M5) &0.007	&3100& 0.998& eR\\
    42& 2MASS J08555655-6146057&93.0& 25.0$\pm$1.2 && (M5) &0.008	&3200& 0.948& eR\\
    44& 2MASS J09280826-6553589&93.7& 23.5$\pm$1.1 && (M5) &0.004	&3100& 0.992& X, eR\\
    45& 2MASS J09522444-6731136&86.2& 22.4$\pm$1.1 && (M5) &0.010	&3100& 0.999& eR\\
    46& 2MASS J09323325-6908439&86.7& 22.2$\pm$1.1 && (M5) &0.010	&3200& 0.998&  eR\\
    49& 2MASS J10222892-6115485&90.1& 23.0$\pm$1.2 && (M5) &0.004	&3100& 0.966& eR\\
    51& UCAC4 116-024312&95.5& 23.1$\pm$1.1 && (M5) &0.011	&3100& 0.951& eR\\
    52& 2MASS J08383351-6716368&85.0& 23.2$\pm$1.1 && (M5) &0.003	&3000& 0.992& eR\\
    57& 2MASS J09430009-6422290&95.4& 23.8$\pm$1.1 && (M5) &0.009	&3100& 0.970& X, eR\\
    59& UCAC4 120-028626&82.4& 22.1$\pm$1.1 && (M5) &0.013	&3200& 0.998& eR\\
    60& UCAC4 120-021739&74.1& 22.6$\pm$1.1 && (M5) &0.008&	3200& 0.970& R, eR\\
    63& 2MASS J10082205-6757040$^a$&86.6& 21.7$\pm$1.1 && (M5) &0.007&	3100& 0.972& eR\\
    64& 2MASS J09133635-6522114&87.3& 23.6$\pm$1.1 && (M5) &0.009&	3200& 0.999& eR\\
    65& UCAC4 096-017311&79.1& 21.9$\pm$1.1 && (M5) &0.011&	3200& 0.967& eR\\
    26& 2MASS J08524155-6401229&82.2& 23.8$\pm$1.1 && (M6) &0.003	&3000& 0.993& X\\
    34& 2MASS J09191188-6640123&73.8& 22.4$\pm$1.1 && (M6) &0.002	&2700& 0.949& uleR\\
    35& 2MASS J09202620-6329547&90.0& 24.1$\pm$1.1 && (M7) &0.001	&2700& 0.997& X\\
    41& 2MASS J09244959-6216319&88.4& 24.0$\pm$1.2 && (M7) &0.001	&2700& 0.994& uleR\\
    50& 2MASS J09313619-6659363&85.0& 22.8$\pm$1.1 && (M8) &0.001&	2350& 0.998& ulX\\ 
    58& Gaia DR3 5246642713478298624&88.8& 23.6$\pm$1.1 && (M8) &0.001&	2400& 0.994& uleR\\
   
  \hline 
    \noalign{\smallskip}
    \multicolumn{3}{l}{Uncertain candidate} &&& & & & \\
    \noalign{\smallskip}
    \hline 
    \noalign{\smallskip}
    & 2MASS J10223677-5848037$^d$&81.9& 22.4$\pm$4.4& {\it a} & & 0.147 & 4050 & 0.546$^*$&R, eR\\
    \hline 
    \noalign{\smallskip}
    \multicolumn{10}{l}{\footnotesize Resolved CPM binaries are indicated in pairs with $^a$, $^b$, $^c$, and $^d$.}\\
    \multicolumn{10}{l}{\footnotesize Spectral types presented in parenthesis were derived by \citetalias{gagne2018a} using a photometric estimation. The remaining spectral types are taken from}\\
    \multicolumn{10}{l}{\footnotesize \cite{houk1975}.}\\
    \multicolumn{10}{l}{\footnotesize Reference for RV values: {\it a}-{\it Gaia} DR3, {\it b}-{\it Gaia} DR2, {\it c}- \cite{kunder2017}, {\it d}-\cite{andersen1983}, {\it e}-\cite{gontcharov2006}. For the }\\
    \multicolumn{10}{l}{\footnotesize candidate members we report the optimal RV from BANYAN $\Sigma$.}\\
    \multicolumn{10}{l}{\footnotesize We report X-ray detections from ROSAT (R), eROSITA (eR), and {\it XMM-Newton} (X), upper limits from {\it XMM-Newton} (ulX), and upper limits} \\
    \multicolumn{10}{l}{\footnotesize from eROSITA (uleR).}
\end{longtable}

\section{Best-fit spectral model}\label{appen:xmmSpectraAll}

Figures~\ref{fig:xmmSpectraApend1} and \ref{fig:xmmSpectraApend2} display the {\it XMM-Newton} spectra of the seven sources fitted using a single-component {\texttt APEC} model, as discussed in Sect.~\ref{subsect:XMMspectra}. The parameters in Table~\ref{table:cstatResults} are obtained from the best-fit model presented here.

\begin{figure*}
\centering
    \includegraphics[width=0.37\textwidth,trim=0.2cm 0.3cm 0.1cm 0.2cm,clip]{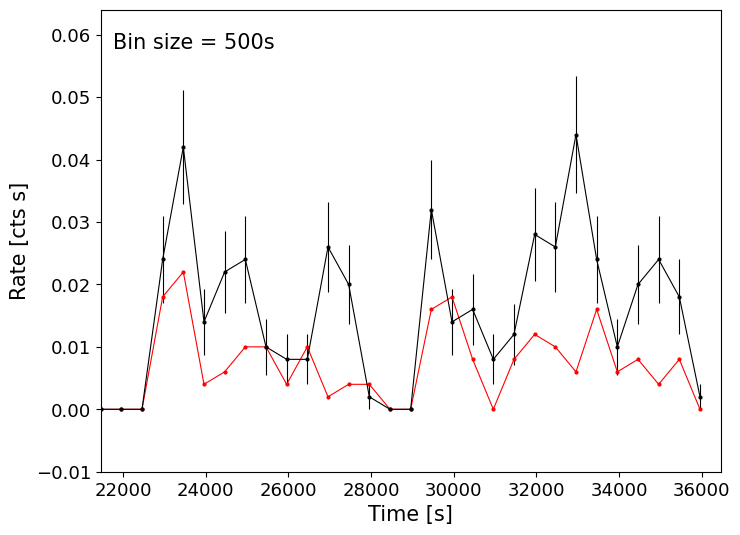}\hfill
    \includegraphics[width=0.36\textwidth,trim=0.2cm 0.2cm 0.1cm 0.97cm,clip]{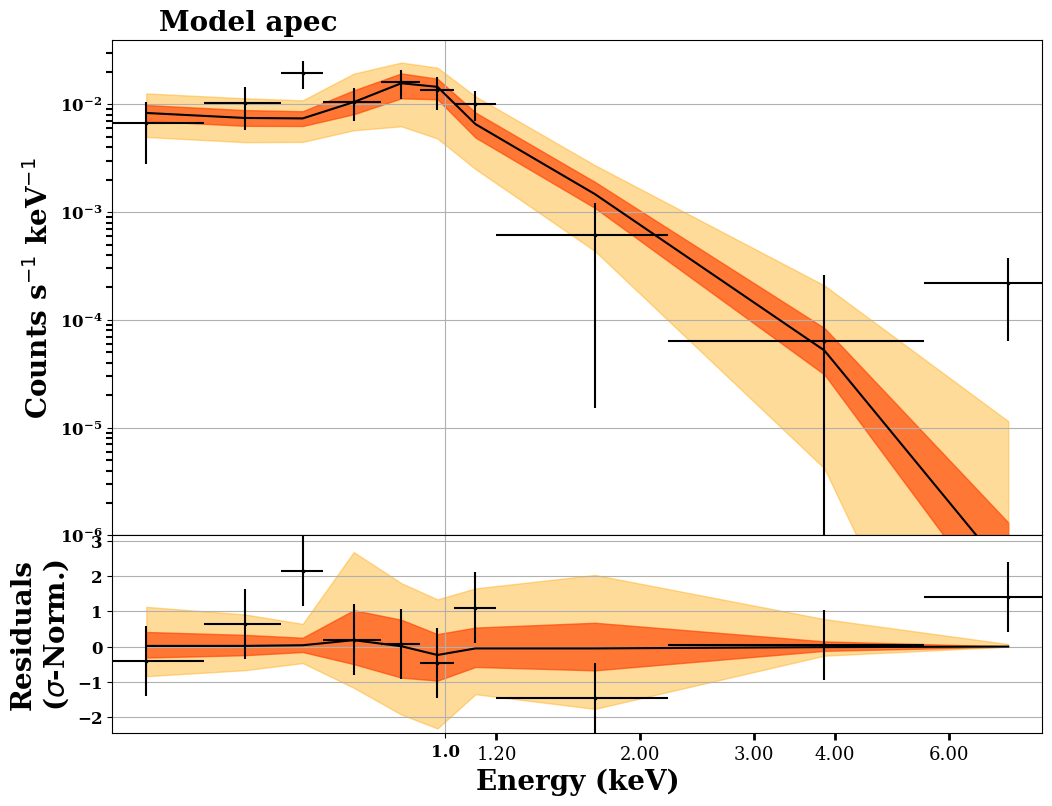}\hfill
    \includegraphics[width=0.26\textwidth,trim=0.3cm 0.2cm 0.2cm 0.2cm,clip]{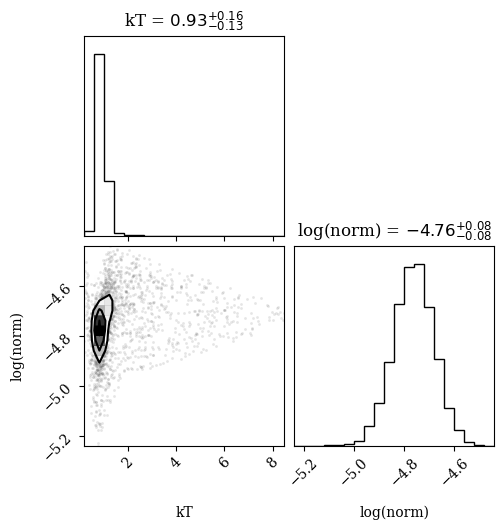}\\
    \includegraphics[width=0.37\textwidth,trim=0.2cm 0.3cm 0.1cm 0.2cm,clip]{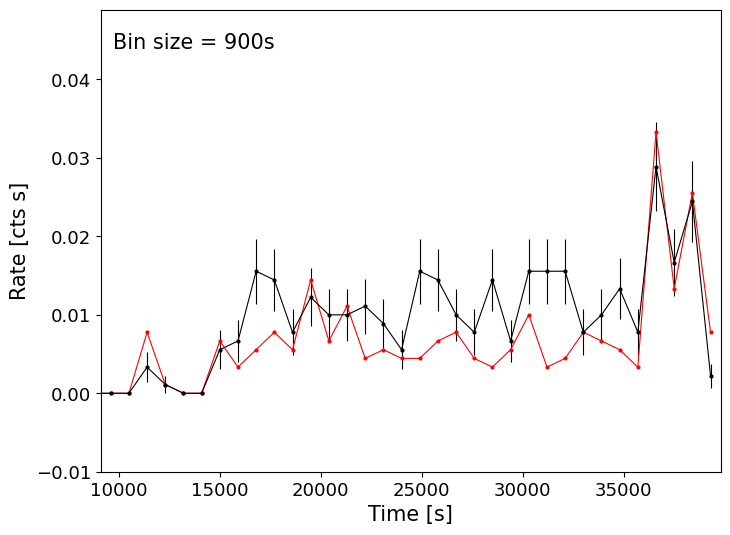}\hfill
    \includegraphics[width=0.36\textwidth,trim=0.2cm 0.2cm 0.1cm 0.97cm,clip]{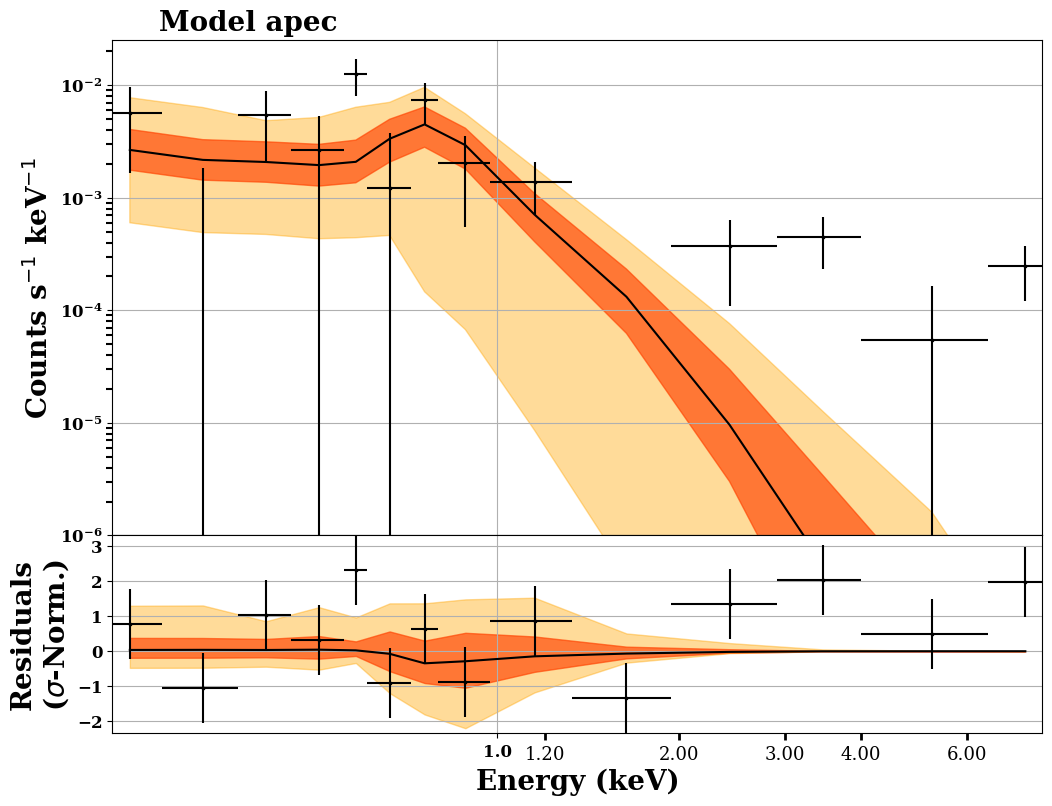}\hfill
    \includegraphics[width=0.26\textwidth,trim=0.3cm 0.2cm 0.2cm 0.2cm,clip]{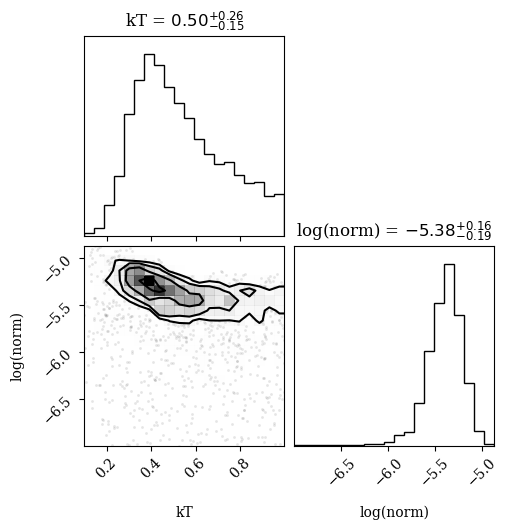}\\
    \includegraphics[width=0.37\textwidth,trim=0.2cm 0.3cm 0.1cm 0.2cm,clip]{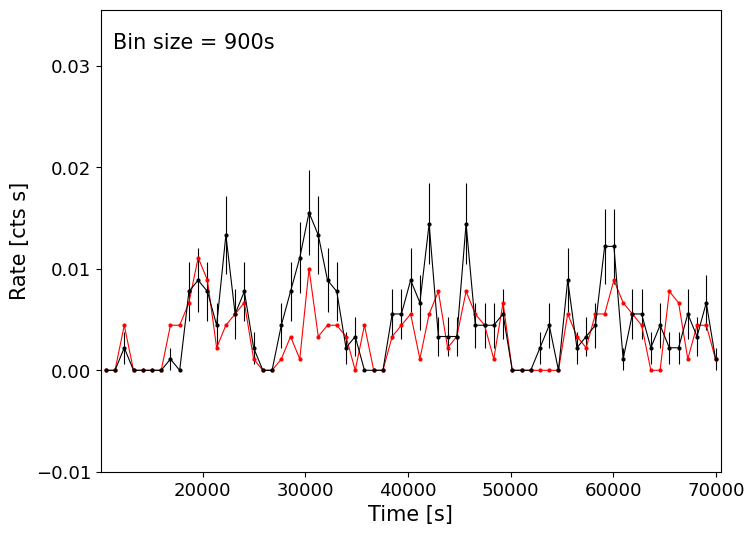}\hfill
    \includegraphics[width=0.36\textwidth,trim=0.2cm 0.2cm 0.1cm 0.97cm,clip]{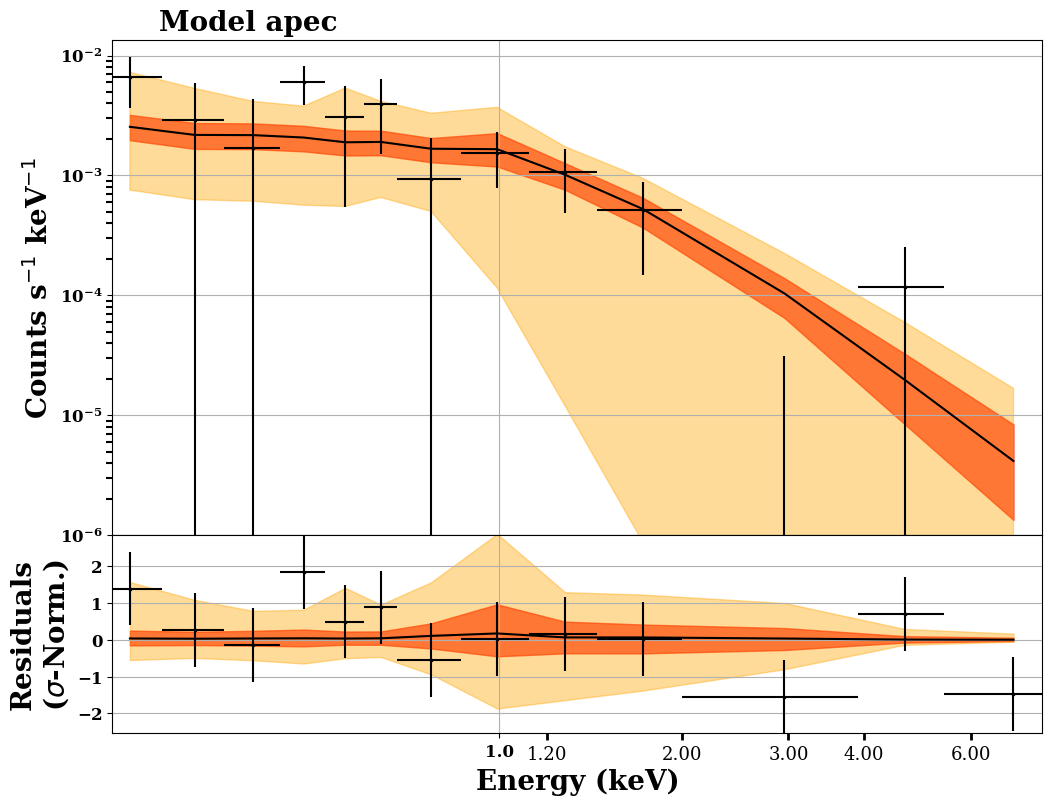}\hfill
    \includegraphics[width=0.26\textwidth,trim=0.3cm 0.2cm 0.2cm 0.2cm,clip]{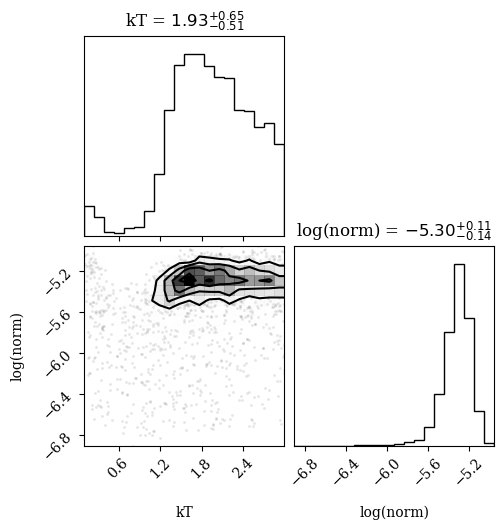}\\
    \includegraphics[width=0.37\textwidth,trim=0.2cm 0.3cm 0.1cm 0.2cm,clip]{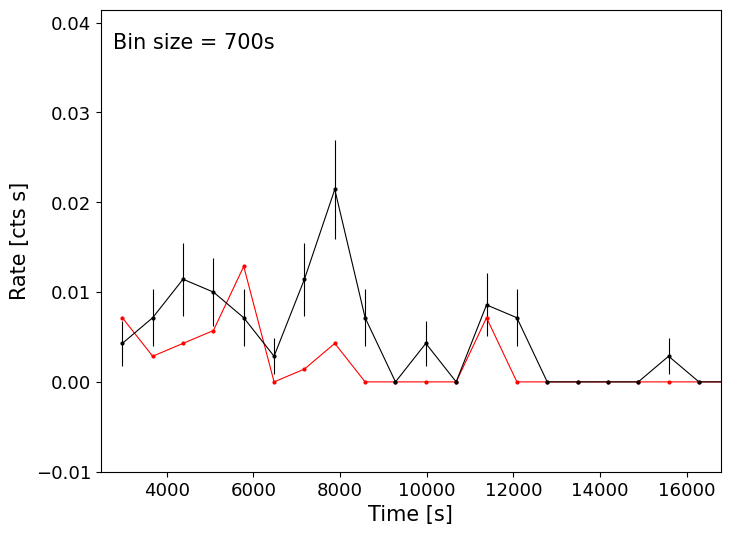}\hfill
    \includegraphics[width=0.36\textwidth,trim=0.2cm 0.2cm 0.1cm 0.97cm,clip]{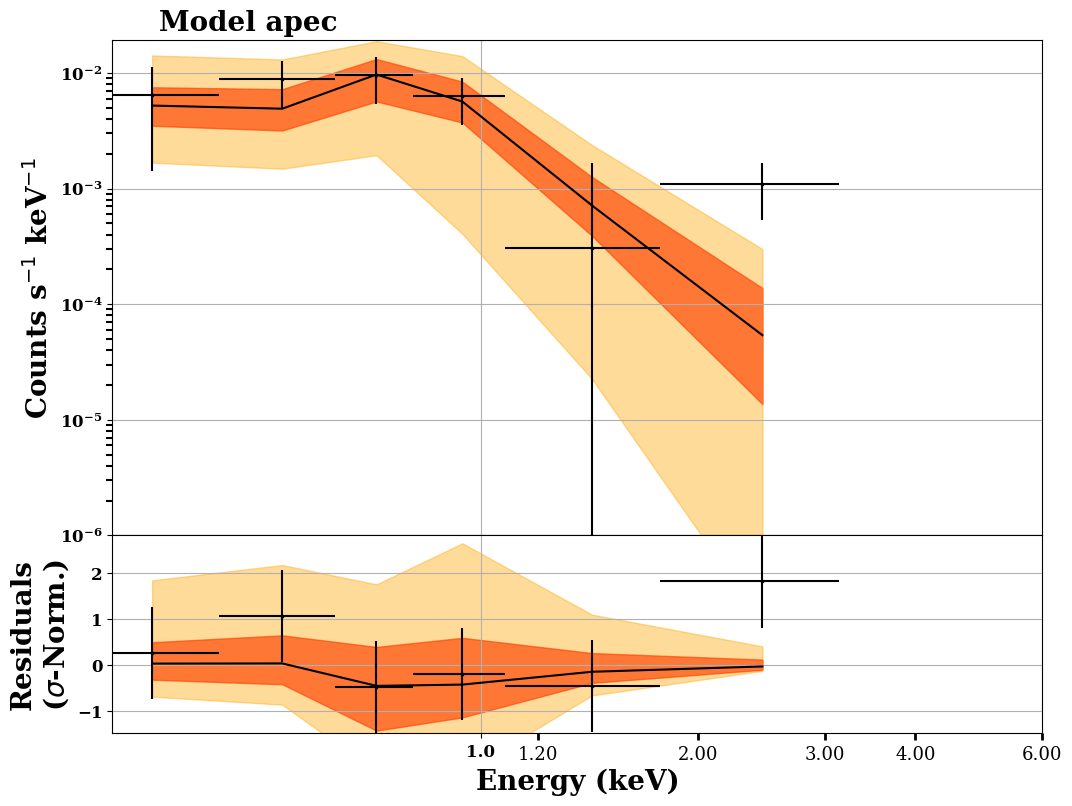}\hfill
    \includegraphics[width=0.26\textwidth,trim=0.3cm 0.2cm 0.2cm 0.2cm,clip]{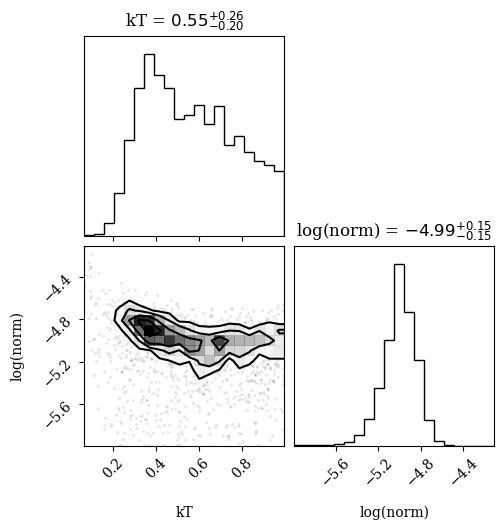}
    \caption{X-ray light curve, spectrum, and best-fit spectral parameters for {\it XMM-Newton} observations of CMs detected with EPIC/pn. {\it Left panels:} EPIC/pn light curve in the {\it XMM-Newton} broadband (0.2 -12 keV) with a bin size given in the legend. The X-ray light curve represents the background-subtracted source signal (black), with the background signal (red) included for comparison. {\it Middle panel:} EPIC/pn spectra with best-fit single-component {\texttt APEC} model, and residuals. The 1$\sigma$ and 3$\sigma$ confidence regions are shown in orange and light orange respectively. {\it Right panel:} Corner plots for the best-fit spectral parameters, showing contours for the 1$\sigma$, 2$\sigma$ and 3$\sigma$ confidence regions. We present from the top to the bottom row the results for: VCA~22, VCA~26, VCA~35, and VCA~38.}
    \label{fig:xmmSpectraApend1}
\end{figure*}

\begin{figure*}
\ContinuedFloat
\centering
    \includegraphics[width=0.37\textwidth,trim=0.2cm 0.3cm 0.1cm 0.2cm,clip]{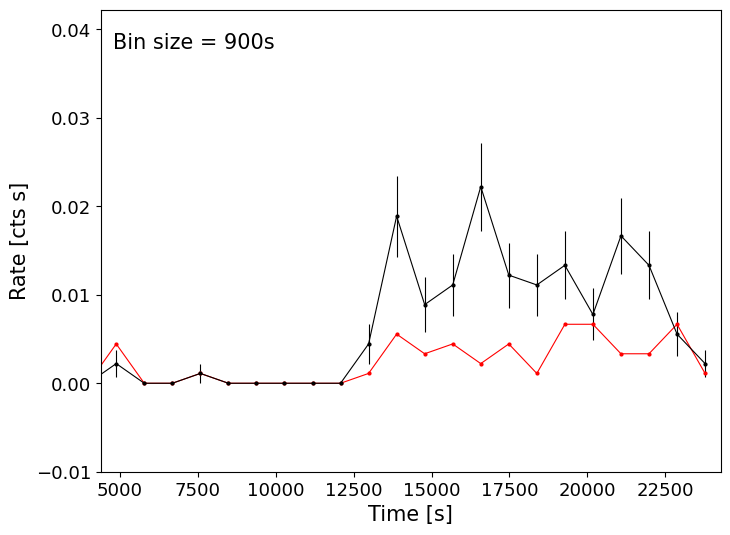}\hfill
    \includegraphics[width=0.36\textwidth,trim=0.2cm 0.2cm 0.1cm 0.97cm,clip]{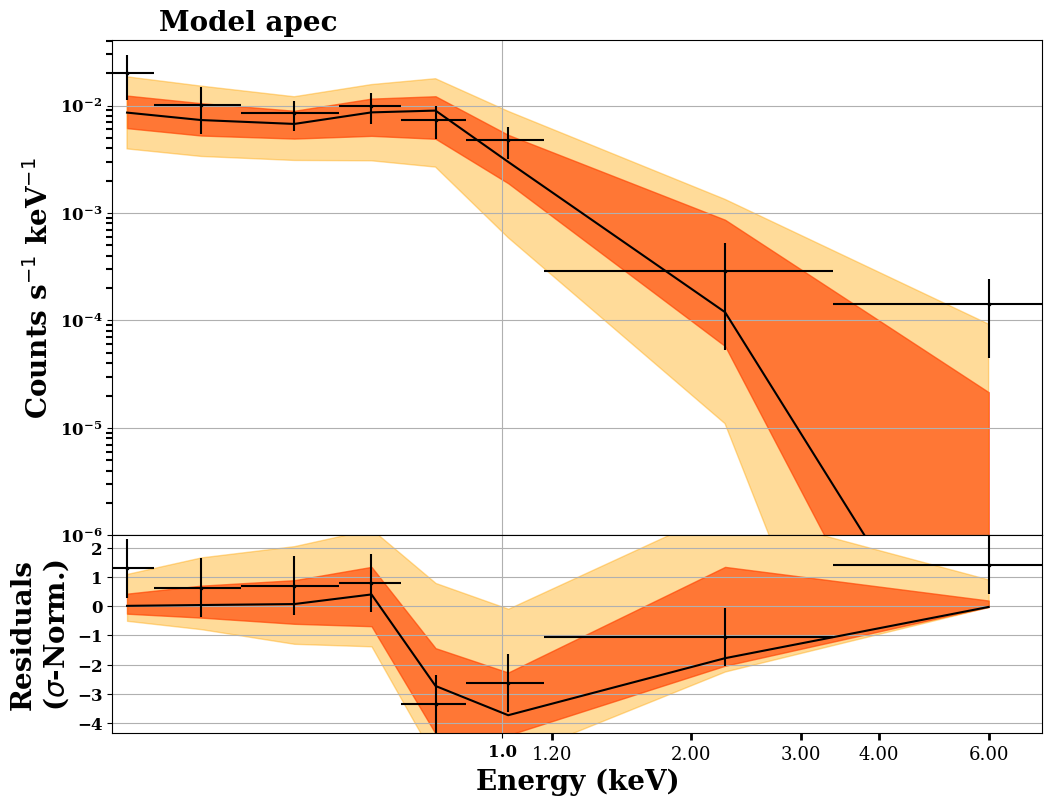}\hfill
    \includegraphics[width=0.26\textwidth,trim=0.3cm 0.2cm 0.2cm 0.2cm,clip]{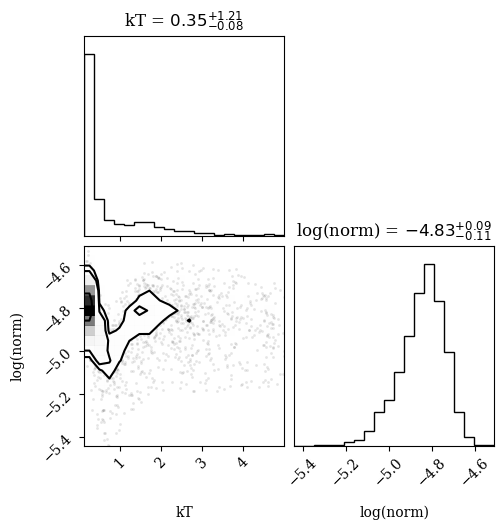}\\
    \includegraphics[width=0.37\textwidth,trim=0.2cm 0.3cm 0.1cm 0.2cm,clip]{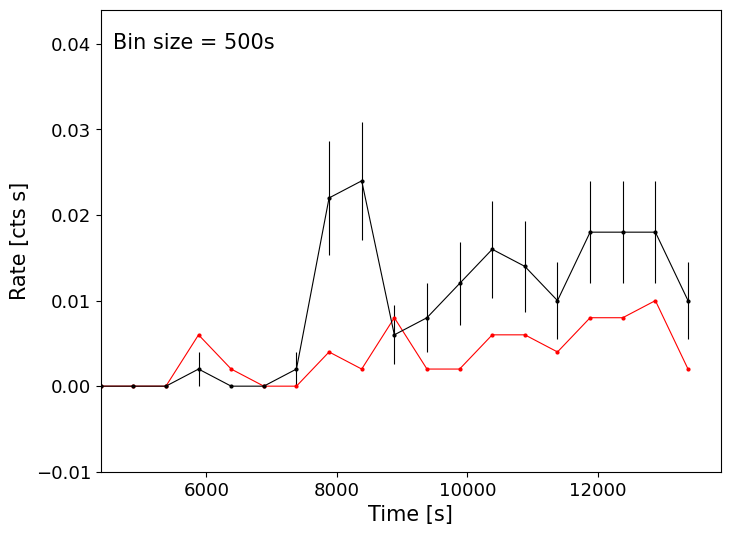}\hfill
    \includegraphics[width=0.36\textwidth,trim=0.2cm 0.2cm 0.1cm 0.97cm,clip]{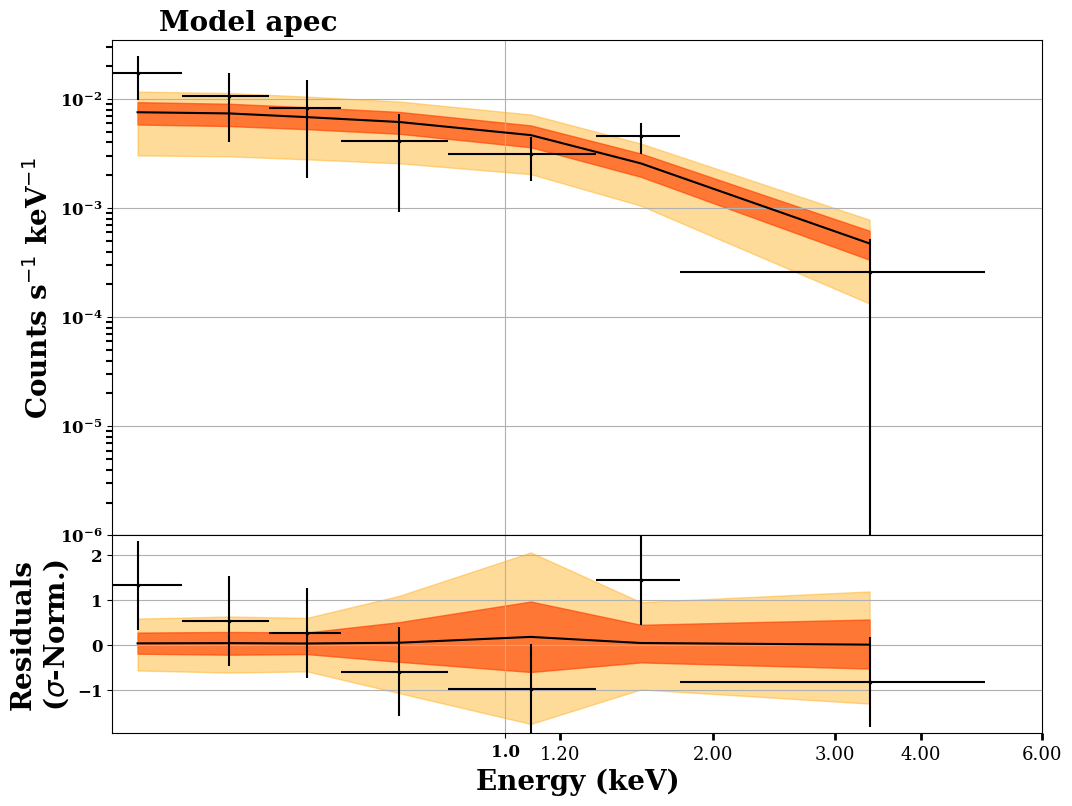}\hfill
    \includegraphics[width=0.26\textwidth,trim=0.3cm 0.2cm 0.2cm 0.2cm,clip]{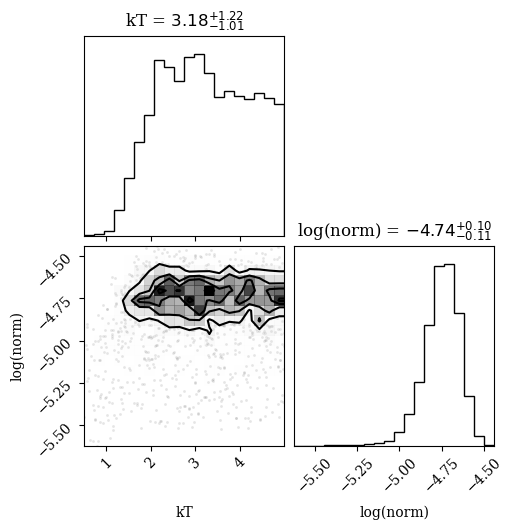}\\
    \includegraphics[width=0.37\textwidth,trim=0.2cm 0.3cm 0.1cm 0.2cm,clip]{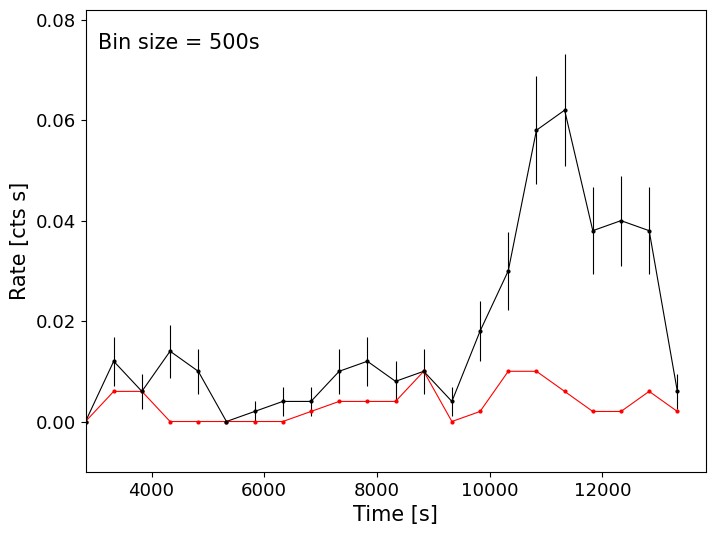}\hfill
    \includegraphics[width=0.36\textwidth,trim=0.2cm 0.2cm 0.1cm 0.97cm,clip]{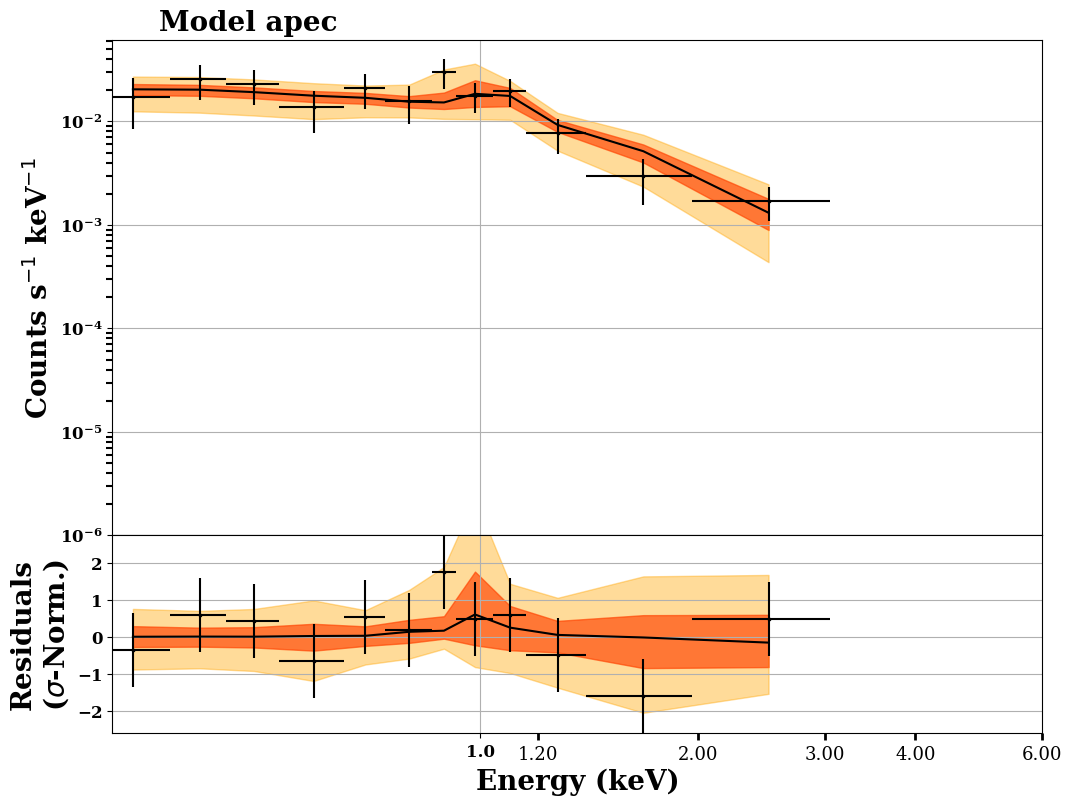}\hfill
    \includegraphics[width=0.26\textwidth,trim=0.3cm 0.2cm 0.2cm 0.2cm,clip]{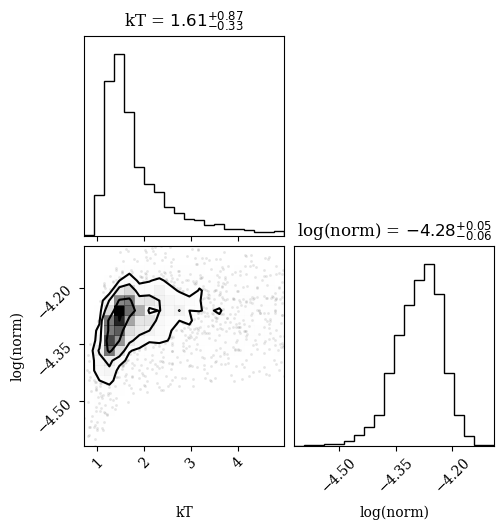}
    \caption{Continued. From the top to the bottom row: VCA~44, VCA~53, and VCA~57.}
    \label{fig:xmmSpectraApend2}
\end{figure*}

\section{eROSITA light curves}\label{appen:eROSITAlc}

We extracted the eROSITA light curves (eSASS version 240410.0.4) for the VCA members and candidates ensuring that the light curves were binned regularly with a bin size of 1 eRODay (corresponding to a revolution of 4\,hr) and that any empty bins were discarded (fractional exposure equals to 0). Following the prescription from \cite{joseph2024}, a quiescent level was determined for each object, allowing the identification of variability in the light curves. Fig.~\ref{fig:eRASSLCApend} shows the eROSITA light curves of the two most X-ray luminous systems in the VCA, both exhibiting clear signs of flaring in at least two eRASS surveys. No other VCA members or candidates present such strong flaring events, which possibly explains the higher average X-ray luminosities of these objects. 

\begin{figure*}
    \centering
    \includegraphics[width=0.49\textwidth,trim=0.4cm 0.1cm 0.3cm 1.05cm,clip]{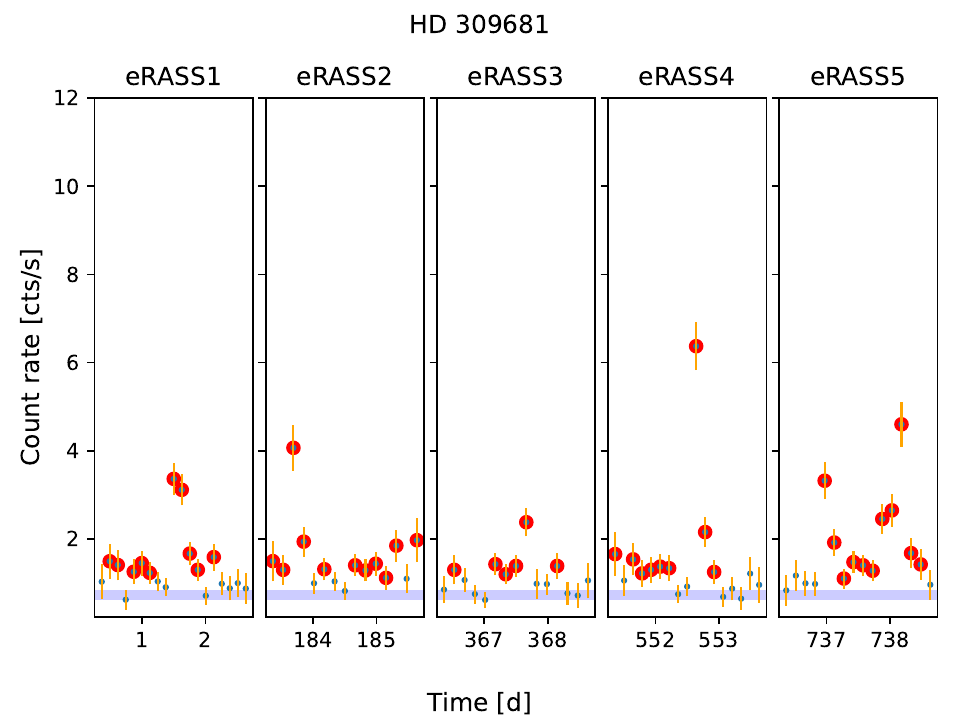}\hfill
    \includegraphics[width=0.49\textwidth,trim=0.4cm 0.1cm 0.3cm 1.05cm,clip]{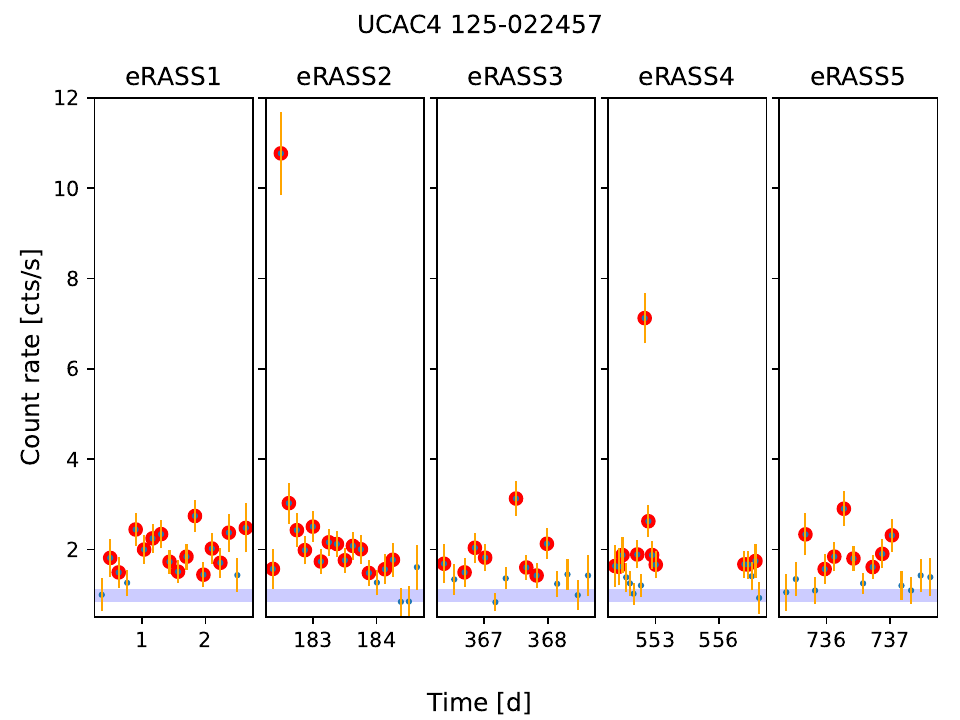}\\
    \caption{eROSITA light curves for VCA stars with elevated X-ray luminosity in Fig~\ref{fig:Tdiff}. The quiescent level is shown as a horizontal band in light purple, with the bins above the quiescent level highlighted in red. {\it Left panel:} VCA~4. {\it Right panel:} Optically resolved CPM binary system composed of VCA~28 and VCA~11.}
    \label{fig:eRASSLCApend}
\end{figure*}

\end{appendix}

\end{document}